\documentclass[a4paper,11pt, table]{article}
\pdfoutput=1

\usepackage{jheppub}

\usepackage{epsf,amsmath,amssymb,graphicx,dcolumn}
\usepackage{scalefnt}
\usepackage{hyperref}
\usepackage{cleveref,etoolbox,color,soul}
\usepackage{listings,booktabs}
\usepackage[utf8]{inputenc}
\usepackage{epsfig}
\usepackage{graphicx}
\usepackage[colorinlistoftodos]{todonotes}
\usepackage{multirow}

\definecolor{linkcolor}{rgb}{0,0,0.5}

\def\Y{\mathbf{Y}}

\begin{document}
\title{The hidden side of scalar-triplet models with spontaneous CP violation}

\author[a,b]{P.M.~Ferreira,}
\author[c,b]{B.L.~Gon\c{c}alves,}
\author[c]{and F.R.~Joaquim}

\affiliation[a]{Instituto Superior de Engenharia de Lisboa, Instituto Polit\'ecnico de Lisboa 1959-007 Lisboa, Portugal}
\affiliation[b]{Centro de F\'isica Te\'orica e Computacional, Faculdade de Ci\^encias,
Universidade de Lisboa, Campo Grande, Edif\'icio C8 1749-016 Lisboa, Portugal}
\affiliation[c]{Departamento de F\'isica and CFTP, Instituto Superior T\'ecnico, Universidade de Lisboa, Lisboa, Portugal}

\emailAdd{pmmferreira@fc.ul.pt}
\emailAdd{bernardo.lopes.goncalves@tecnico.ulisboa.pt}
\emailAdd{filipe.joaquim@tecnico.ulisboa.pt}

\abstract{Scalar triplet extensions of the Standard Model provide an interesting playground for the explanation of neutrino mass suppression through the type-II seesaw mechanism. Propelled by the possible connections with leptonic CP violation, we explore under which conditions spontaneous CP violation can arise in models with extra scalar triplets. The minimal model satisfying such conditions requires adding two such triplets to the SM field content. For this model, the scalar mass spectrum in both the CP-conserving and spontaneous CP-violating scenarios is studied. In the former case, a decoupling limit for the new scalars can be achieved, while this is not the case when CP is spontaneously broken. In particular, we show that the existence of two light neutral scalars with masses below a few tenths of GeVs is unavoidable in the CP-violating case. Using matrix theory theorems, we derive upper bounds for the masses of those light scalars and briefly examine whether they can still be experimentally viable. Other interesting features of the scalar mass spectrum are discussed as, e.g., the existence of relations among the charged and neutral scalar masses.}

\maketitle

\section{Introduction}

With the discovery of the Higgs boson by the ATLAS and CMS Collaborations~\cite{Aad:2012tfa,Chatrchyan:2012xdj} at the Large Hadron Collider (LHC), the particle content of the Standard Model (SM) has been confirmed. Though the SM is in remarkable agreement with most experimental data delivered so far, there are still important questions which remain unanswered within the standard theory: among others, the origin of fermion masses; the observed matter--antimatter asymmetry; the existence of dark matter; and even the recent confirmation of a 4.2$\sigma$ discrepancy on the muon anomalous magnetic moment~\cite{Muong-2:2021ojo,Muong-2:2006rrc}. There is a plethora of SM extensions~\footnote{In this work we will only consider renormalisable extensions of the SM. Different conclusions, of course, could be reached
if one were to consider effective interactions stemming from operators of dimensions higher than four.} which provide answers for one or several of these questions, some of them relying on enlarging the SM field content with extra spin-zero fields. Popular models include scalar gauge singlets~\cite{McDonald:1993ex,Barger:2008jx}, or a second scalar doublet, yielding the two Higgs Doublet Model (2HDM), first introduced by T.~D.~Lee in 1973~\cite{Lee:1973iz}. In both cases, spontaneous CP violation (SCPV) may be realized, suitable dark-matter candidates are provided, and the dynamics of the first-order electroweak phase transition may be altered~\cite{Turok:1990,Turok:1991,McDonald:1994,Cline:1996,Bohdan:2018}. SCPV is appealing for several reasons, one of which is the possibility of explaining the matter-antimatter
asymmetry via a mechanism of spontaneous symmetry breaking, thus with the same kind of mechanism which explains
the fact that the electroweak gauge bosons are massive.
Another scenario much studied in the literature is the scalar-triplet model where the usual hypercharge $Y = 1$ SM Higgs doublet is complemented with a scalar SU(2) triplet with hypercharge $Y = 2$~\cite{Konetschny:1977bn}. The main motivation for considering such possibility is that small neutrino masses may be generated via the type-II seesaw mechanism~\cite{Mohapatra:1979ia,Magg:1980ut,Cheng:1980qt,Schechter:1980gr,Lazarides:1980nt}. Although some versions of the model can also accommodate dark matter candidates~\cite{Kanemura:2012rj},
SCPV is not possible in the case with only one triplet.
The possibility of dynamical CP violation triggered by complex vacuum expectation values (VEVs) of neutral scalar-triplet components is interesting in the sense that such CP-violating effects can, in principle, be communicated to the lepton sector via Yukawa interactions. This would, for instance, lead to a connection between Dirac and Majorana leptonic CP violation (for a review see Ref.~\cite{Branco:2011zb}) and the fundamental properties of the vacuum~\cite{Branco:2012vs}. In more elaborated scenarios, relations can also be established with the problem of the matter-antimatter asymmetry in the Universe in the context of leptogenesis induced by heavy triplet decays into leptons~\cite{Ma:1998dx,Felipe:2013kk}.

SM extensions with scalar triplets have a much richer spectrum, including new neutral, charged and doubly-charged scalars, opening up the possibility for new experimental searches. On the other hand, any multiscalar model must account for the fact that the LHC has established that the 125 GeV Higgs boson has properties similar, within current experimental uncertainties, to those expected for the SM Higgs particle (see for instance~\cite{Aad:2015zhl,Khachatryan:2016vau}), and also for the non-observation, thus far, of other scalars. These requirements can be met in several ways -- for instance, in the inert 2HDM there exists a scalar with tree-level couplings identical to those of the SM Higgs boson, the extra scalar particles being inert, i.e. they do not couple to fermions or gauge bosons at tree level. This feature stems from imposing a discrete $Z_2$ symmetry which is unbroken by the vacuum, guaranteeing the existence of a suitable dark matter candidate. Another possibility is to choose (though without fine tuning) the parameters of a certain model such that the extra scalars are very heavy. For instance, within the 2HDM, where sum rules for the scalar-gauge boson couplings exist, such regime could explain in a fairly natural way why the heavier CP-even state would have escaped detection at the LHC in searches for resonances in $Z$ boson pairs. This {\em decoupling limit} is achieved due to one of the quadratic coefficients of the scalar potential being unconstrained by the minimisation conditions which determine the CP-conserving vacuum of the theory. However, it has been shown that a 2HDM with a CP-violating minimum does not exhibit a decoupling limit~\cite{Nebot:2020niz}. In that case, there is an upper bound for the extra scalar masses resulting from unitarity and perturbativity constraints which limit the magnitude of the quartic couplings.

In this work, we investigate whether SCPV violation may occur in models with extra scalar triplets. One possible candidate would be the Georgi-Machacek model~\cite{Georgi:1985nv}, which includes a real and a complex triplet with hypercharges $Y=1$ and $Y=2$, respectively. Although much work has been done in this framework~\cite{Chanowitz:1985ug,Gunion:1989ci, Gunion:1990dt, Cheung:1994rp, Aoki:2007ah, Logan:2010en,Chiang:2012cn, Englert:2013wga, Chiang:2013rua, Kanemura:2014bqa, Hartling:2014zca, Chiang:2014bia, Degrande:2015xnm, Chang:2017niy, Keeshan:2018ypw, Ghosh:2019qie,Chiang:2018cgb, Ismail:2020zoz,Hartling:2014xma}, SCPV violation is not possible in it. Motivated by the possible connection with leptonic CP violation, in this work we consider instead the SM extended with two $Y=2$ complex triplets which we dub as the two-scalar-triplet model (2STM). It turns out that this is the minimal triplet model~\footnote{By ``minimal" we refer to the field content of the model, not its number of independent parameters.} which meets the conditions for SCPV to occur and lead to interesting leptonic CP-violating effects which are being probed, for instance, at neutrino oscillation experiments. Moreover, we will see that the 2STM reveals itself as an instructive example of how requiring that the vacuum breaks CP can restrain the decoupling of extra scalar particles by introducing novel features, namely the presence of very light scalars. It also shows how decisive the mass spectrum analysis can be for model-building scenarios with extended scalar sectors.

This paper is organized as follows: in Section~\ref{sec:scpv}, after recapping the main features of the SM extended with one scalar triplet, we introduce the 2STM. The possibility of SCPV is considered and we show that it cannot occur with a single triplet field, but is possible when two triplets are added. For the sake of simplicity, we consider the 2STM endowed with a (softly-broken) continuous global symmetry, which turns out to be the minimal model with two triplets where SCPV can be realised. In Section~\ref{sec:mass} we study in detail the scalar mass spectrum of the 2STM for both CP-conserving and CP-violating minima of the scalar potential. By using matrix-theory theorems we obtain some upper bounds on the scalar masses and investigate whether decoupling is possible for both types of minima. We conclude that the model is non-decoupling  when SCPV occurs. These results are then confirmed by a numerical analysis presented in Section~\ref{sec:num}, where several experimental constraints are considered. Finally, in Section~\ref{sec:conclusions}, we draw our final conclusions.

\section{SCPV in the two-scalar-triplet model}
\label{sec:scpv}

Before introducing the 2STM, which will be the main focus of this work, it is instructive to revisit the most relevant aspects of the one-triplet model, sometimes referred to as the scalar-triplet model (STM). Its scalar sector contains the hypercharge one ($Y=1$) SM Higgs doublet $\Phi$, and a $Y=2$ scalar triplet $\Delta$. The field components of the scalar multiplets $\Phi$ and $\Delta$ are denoted with the usual notation,
\begin{equation}
 \Phi  =
\begin{pmatrix}
\phi^{+} \\
\phi^{0}
\end{pmatrix}
\,\
,
\,\
\Delta =
\begin{pmatrix}
\delta^{+} / \sqrt{2} & \delta^{++}\\
\delta^{0} & - \delta^{+} / \sqrt{2}
\end{pmatrix}
\,,
\label{eq:fields}
\end{equation}
where the $\text{SU}(2)$ matrix representation for $\Delta$ has been considered. With this scalar field content, the most general renormalizable SU(2)$_L\times$U(1)$_Y$ gauge-invariant scalar potential is given by
\begin{equation}
    \begin{split}
V_{S} =& \, m^2 \Phi^{\dagger} \Phi + M^2 \text{Tr}(\Delta^{\dagger} \Delta) + (\mu\Phi^{T} i \tau_2 \Delta^{\dagger} \Phi + \text{H.c.}) \\
+& \, \lambda_1 (\Phi^{\dagger} \Phi)^2 + \lambda [\text{Tr}(\Delta^{\dagger} \Delta)]^2 + \tilde{\lambda}\text{Tr}[(\Delta^{\dagger} \Delta)^2] + \lambda^\prime (\Phi^{\dagger} \Phi) \text{Tr}(\Delta^{\dagger} \Delta) + \hat{\lambda} \Phi^{\dagger} \Delta \Delta^{\dagger}  \Phi \, .
\end{split}
\end{equation}
Note that the trilinear $\mu$ term softly breaks a global $\text{U}(1)$ transformation of the form $\Phi \rightarrow e^{i \alpha}\Phi$ and $\Delta \rightarrow e^{i \alpha'}\Delta$. It is straightforward to see that, without loss of generality, the $\mu$ coefficient can always be made real via rephasings of the doublet and triplet scalars. Electroweak symmetry breaking (EWSB) occurs if the parameters of the potential are such that the neutral components of $\Phi$ and $\Delta$ acquire vacuum expectation values (VEVs), which we represent as
\begin{align}
\langle \phi^{0} \rangle=\frac{v}{\sqrt{2}} \,,\, \langle \delta^{0}\rangle=\frac{u e^{i\theta}}{\sqrt{2}} \,,\,v^2+2u^2\simeq(246 \, \text{GeV})^2 \,,
\label{eq:VEVs1T}
\end{align}
where we have used the freedom to redefine $\Phi$ such that $\langle \phi^{0} \rangle$ is real. The last relation above ensures consistent EWSB, i.e. this vacuum yields correct values for the masses of the electroweak gauge bosons. Notice, however, that unlike  models with an arbitrary number of Higgs  doublets, in the triplet model the electroweak precision parameter $\rho$ deviates from unity already at tree-level. As such, compatibility with electroweak precision data forces the triplet VEV $u$ to be, at most, $8 \, \text{GeV}$~\cite{Arhrib:2011uy,Arhrib:2011vc,Kanemura:2012rj,Aoki:2012yt,Aoki:2012jj}. We mention in passing that the STM can have charge-breaking (CB) vacua, some of which can coexist with minima of the type of Eq.~\eqref{eq:VEVs1T}. In fact, the minimization equations of the STM allow for simultaneous solutions which may break different symmetries. As was shown  in~\cite{Ferreira:2019hfk}, charge-preserving and CB vacua are found to, for certain regions of parameter space, coexist. And indeed, it is even possible that a CB minimum, deeper than the SM-like one, exist. An analytical and numerical analysis of the STM vacuum structure was performed in~\cite{Ferreira:2019hfk}.

Regarding the scalar-fermion couplings, the doublet $\Phi$ couples to fermions in exactly the same way as in the SM, while the triplet $\Delta$ can couple to the SM lepton doublets $L_\alpha$ as
\begin{equation}
 \mathcal{L}_{\rm Yuk} = \Y_{\alpha\beta}^\Delta \,L_\alpha^T \mathcal{C}^\dag  i \tau_2 \Delta \,L_\beta+
 {\rm H.c.}\,,\,
L_\alpha =
\begin{pmatrix}
\nu_L \\
\ell_L
\end{pmatrix}_\alpha\,,\, (\alpha,\beta)=e,\mu,\tau\,.
\label{eq:Lyuk1}
\end{equation}
Here, $\mathcal{C}$ is the charge-conjugation matrix, $\tau_2$ is the complex Pauli matrix and $\Y^\Delta$ is a general $3\times 3$ complex (symmetric) matrix. The triplet VEV $u$, which is induced upon EWSB, generates a neutrino Majorana mass matrix $m_\nu= \Y_\Delta u/2$. Small neutrino masses may result from either suppressed Yukawa couplings and/or a tiny $u$ VEV.
In the latter case, and considering $\mathcal{O}(1)$ Yukawa couplings, one may get $m_\nu \sim u\sim \mu v^2/M^2 \sim {\rm 0.1~eV}$ by considering $\mu v/ M^2 \sim 10^{-13}$ which, for $\mu \sim v$, implies $M \sim 10^{9}$~GeV. However, if $\mu \ll v$ (which would be natural in the t'Hooft sense in the context of a broken softly-broken lepton number symmetry), the scale $M$ can be lowered to e.g. the electroweak scale.

Let us now investigate the possibility of SCPV in the STM. As already mentioned, it is always possible to render the $\mu$ parameter real through field redefinitions, making $V_S$ real and the Lagrangian CP conserving. Then, the non-trivial extremisation conditions are:
\begin{equation}
\begin{aligned}
v \left[ 2 m^2 + 2 \lambda_1 v^2 + (\lambda' + \hat{\lambda}) v^2 - 2 \sqrt{2} \mu u \cos{\theta} \right]&=0 \, , \\
    -\sqrt{2}\mu v^2 + u \left[ 2M^2 + (\lambda' + \hat{\lambda}) v^2 + 2 (\lambda + \tilde{\lambda}) u^2 \right] \cos{\theta}&=0 \, , \\
    \mu \, v u \sin{\theta} &= 0\, .
\label{eq:last_one_triplet}
\end{aligned}
\end{equation}
It is straightforward to see that, for the cases of interest with $\mu,v,u\neq 0$, the only possible solution for the last equation above is $\theta=n\pi$. Thus, CP cannot be spontaneously broken in a model with a single $\Delta$ and, therefore, leptonic CP violation can only be accounted for with complex Yukawa couplings. With these minimization conditions, we find that the STM has a richer scalar spectrum than the SM -- it includes two CP-even scalars, $h$ (the 125~GeV SM-like Higgs boson) and $H$, a CP-odd one $A$, a charged scalar $H^\pm$ and a doubly-charged, $H^{\pm\pm}$. The $h$ and $H$ masses are the eigenvalues of the matrix
\begin{equation}
\mathcal{M}\,=\,\left(\begin{array}{cc}
2\lambda_1 v^2 & -\dfrac{2 u}{v}\,M_{\Delta}^2 +
 (\lambda^\prime + \hat{\lambda})v u \\
 -\dfrac{2 u}{v}\,M_{\Delta}^2 +  (\lambda^\prime + \hat{\lambda})v u &
 M_{\Delta}^2 + 2 (\lambda + \tilde{\lambda}) u^2
\end{array}\right)\,,
\label{eq:mneu}
\end{equation}
while for $A$, $H^\pm$ and $H^{\pm\pm}$ we have
\begin{align}
  m^2_{A} & = M_{\Delta}^2 \left( 1+\dfrac{4 u^{2}}{v^{2}} \right)\;\;,\;\;
  m^2_+  = \left( M_{\Delta}^2 - \dfrac{\hat{\lambda}}{4}v^{2} \right) \left(
  1+\dfrac{2u^{2}}{v^{2}} \right)\;\;,\;\;
  m^2_{++} = M_{\Delta}^{2}-u^{2}\tilde{\lambda}-\dfrac{\hat{\lambda}}{2}v^{2}\,,
  \label{eq:mchch1}
\end{align}
where we have defined $M_\Delta^2$
\begin{equation}
M_{\Delta}^{2}\,\equiv\,\frac{v^{2}\,\mu}{\sqrt{2}\,u}\,.
\label{eq:MD}
\end{equation}
It is then clear that, except for $h$, all scalars may have masses much larger than $m_h$ if $M_{\Delta}$ is very large. It is also possible to choose a region of parameter space such that the matrix of Eq.~\eqref{eq:mneu} is close to  diagonal -- this can be achieved by taking $|\mu| \ll v$ and remembering that $u \ll v$, so that one obtains $u M_{\Delta}^{2}/v \ll v^2$. Thus, in this situation, there is almost no mixing between the doublet and the triplet, and the $h$ scalar has couplings to gauge bosons and fermions almost identical to those of the SM Higgs. At the same time, since the extra scalars stem from the triplet, their interactions with gauge bosons and fermions are heavily suppressed~\cite{Arhrib:2011uy,Kanemura:2012rj}~\footnote{This is obvious for the quark Yukawa interactions since only the doublet couples to quarks. For the leptons, the triplet Yukawa interactions of Eq.~\eqref{eq:Lyuk1} induce interactions between the extra scalars and leptons, but in this limit those too are suppressed.}. Thus the STM is found to have a {\em decoupling limit} for this region of parameter space.

We will now investigate whether the aforementioned properties  of the STM still hold if one introduces a second triplet field. In particular, we wish to investigate the possibility of SCPV. In the 2STM the scalar sector includes the usual doublet $\Phi$, as well as two $Y=2$ triplets $\Delta_{1,2}$ which, following the conventions in \eqref{eq:fields}, we represent by
\begin{equation}
\Delta_{1,2} =
\begin{pmatrix}
\delta^+_{1,2} / \sqrt{2} & \delta^{++}_{1,2}\\
\delta^0_{1,2} & - \delta^+_{1,2} / \sqrt{2}
\end{pmatrix}
\,.
\label{eq:fields12}
\end{equation}
The CP-conserving scalar potential invariant under the SU(2)$_L\times$U(1)$_L$ gauge symmetry and the global U(1) symmetry $\Phi \rightarrow e^{i \alpha}\Phi$ and $\Delta_{1,2} \rightarrow e^{i \alpha_{1,2}} \Delta_{1,2}$ is
\begin{equation}
    \begin{split}
        V_{\rm U(1)} =& \, m^2 \Phi^{\dagger} \Phi + M_{11}^2 \text{Tr}(\Delta_1^{\dagger} \Delta_1) + M_{22}^2 \text{Tr}(\Delta_2^{\dagger} \Delta_2) + \lambda_0 (\Phi^{\dagger} \Phi)^2\\
        &+ {\lambda}_{1} [\text{Tr}(\Delta_1^{\dagger} \Delta_1)]^2
        + {\lambda}_{2} [\text{Tr}(\Delta_2^{\dagger} \Delta_2)]^2
        + {\lambda}_{21} \text{Tr}(\Delta_2^{\dagger} \Delta_2)\text{Tr}(\Delta_1^{\dagger} \Delta_1)
        + {\lambda}_{12} \text{Tr}(\Delta_1^{\dagger} \Delta_2)\text{Tr}(\Delta_2^{\dagger} \Delta_1)\\
       &+ {\tilde{\lambda}}_{1} \text{Tr}[(\Delta_1^{\dagger} \Delta_1)^2]
        + {\tilde{\lambda}}_{2} \text{Tr}[(\Delta_2^{\dagger} \Delta_2)^2]
        + {\tilde{\lambda}}_{21}  \text{Tr}( \Delta_2^{\dagger} \Delta_2\Delta_1^{\dagger} \Delta_1)
        + {\tilde{\lambda}}_{12}  \text{Tr}(\Delta_1^{\dagger} \Delta_2 \Delta_2^{\dagger} \Delta_1) \\
        &+ {\lambda}_{1}^{\prime} \text{Tr}(\Delta_1^{\dagger} \Delta_1) \Phi^{\dagger} \Phi
        + {\lambda}_{2}^{\prime} \text{Tr}(\Delta_2^{\dagger} \Delta_2)\Phi^{\dagger} \Phi +{\hat{\lambda}}_{1} \Phi^{\dagger} \Delta_1 \Delta_1^{\dagger} \Phi
        + {\hat{\lambda}}_{2} \Phi^{\dagger} \Delta_2 \Delta_2^{\dagger} \Phi
    \end{split}
\label{eq:Vs}
\end{equation}
where all parameters are real. To avoid unwanted massless Goldstone bosons arising upon spontaneous breaking of the exact global U(1) symmetry, we introduce the most general soft symmetry-breaking potential
\begin{align}
V_{\rm SB}=  M_{12}^2 [ \text{Tr}(\Delta_1^{\dagger} \Delta_2) + \text{Tr}(\Delta_2^{\dagger} \Delta_1) ]+(\mu_1 \Phi^{T} \mathrm{i} \tau_2 \Delta_1^{\dagger} \Phi + \mu_2 \Phi^{T} \mathrm{i} \tau_2 \Delta_2^{\dagger} \Phi+  \text{H.c.})\,,
\label{eq:SB12}
\end{align}
with real $M_{1,2}^2$ and $\mu_{1,2}$, to ensure CP invariance. Bear in mind that it would be possible to write a fully symmetric potential with only part of those soft-breaking terms: for instance, one could set $M_{12} = 0$ and $\mu_2 = 0$ by considering the symmetry $\Phi\rightarrow i \Phi$ and $\Delta_{1} \rightarrow - \Delta_{1}$.

As discussed above, in the STM it is not possible to obtain VEVs with a relative complex phase. What about in the 2STM? Given that a complex VEV for $\Phi$ can always be made real through a phase redefinition of $\Delta_{1,2}$, we take
\begin{equation}
\langle \Phi \rangle = \frac{1}{\sqrt{2}}
\begin{pmatrix}
0 \\
v
\end{pmatrix}
\,\
,
\,\
\langle \Delta_1 \rangle= \frac{1}{\sqrt{2}}
\begin{pmatrix}
0 & 0\\
u_1 e^{i\theta_1} & 0
\end{pmatrix}
\,\
,
\,\
\langle \Delta_2 \rangle = \frac{1}{\sqrt{2}}
\begin{pmatrix}
0 & 0\\
u_2 e^{i\theta_2} & 0
\end{pmatrix}\,,
\label{eq:cpviolating}
\end{equation}
being $v$, $u_1$ and $u_2$ real. We must still guarantee that $v^2+2u^2\simeq(246 \, \text{GeV})^2$, with $u^2=u_1^2+u_2^2$, to ensure viable EWSB and, as before, $u$ must be below about 8 GeV to comply with the $\rho$ parameter constraints. Let us now define $\tan\beta = u_2/u_1$, so that  $u_1 = u\, c_\beta$ and $u_2 = u \,s_\beta$, where we use the simplified notation $c_x \equiv \cos{x}$, $s_x \equiv \sin{x}$.
Eq.~\eqref{eq:cpviolating} is the most general VEV configuration for a neutral vacuum. To obtain the VEV values, we perform the derivatives of the potential in Eqs.~\eqref{eq:Vs} and \eqref{eq:SB12}, with respect to all scalar field components. Given the neutral VEVs considered above, many such derivatives yield trivial ($0 = 0$) conditions. In the end, the minimization  of the scalar potential leads to five necessary and sufficient non-trivial conditions for the existence of a vacuum solution such as that of Eq.~\eqref{eq:cpviolating}\footnote{These five equations are, however, not sufficient to guarantee that this solution is unique, or that it corresponds to a minimum of the potential. To verify the latter, one would need to compute the scalar mass matrices of the model.}:
\begin{equation}
\begin{aligned}
        v \left[ 2 m^2 + 2 \lambda_0 v^2 + (\Lambda_1 c_\beta^2 + \Lambda_2 s_\beta^2) u^2 - 2 \sqrt{2} u (\mu_1
        c_{\beta} c_{\theta_1} + \mu_2 s_{\beta} c_{\theta_2} ) \right] &= 0 \, ,  \\
        u c_\beta c_{\theta_1} \left[ 2 M_{11}^2 + \Lambda_1 v^2 +  (2 \Lambda_3 c_\beta^2 + \Lambda_4 s_\beta^2) u^2
        \right] + 2 M_{12}^2 u s_\beta c_{\theta_2} - \sqrt{2} \mu_1 v^2 &= 0  \, , \\
        u s_\beta c_{\theta_2} \left[ 2 M_{22}^2 + \Lambda_2 v^2 + (\Lambda_4 \, c_\beta^2 + 2\Lambda_5 s_\beta^2) u^2
        \right] + 2 M_{12}^2 u c_\beta c_{\theta_1} - \sqrt{2} \mu_2 v^2 &= 0 \, ,  \\
       2 M_{12}^2 u s_\beta s_{\theta_2} + u c_\beta s_{\theta_1} \left[ 2 M_{11}^2 + \Lambda_1 v^2 +  (2\Lambda_3
       c_\beta^2 + \Lambda_4 s_\beta^2) u^2 \right] &= 0  \,,\\
        u v \left( \mu_1 c_\beta s_{\theta_1} + \mu_2 s_\beta s_{\theta_2} \right) &= 0 \, ,
          \label{eq:last_two_triplet}
\end{aligned}
\end{equation}
where the $\Lambda_i$ coefficients are defined in terms of the dimensionless parameters in $V_{\rm U(1)}$ as,
\begin{equation}
    \Lambda_1 = \lambda^{\prime}_{1} + \hat{\lambda}_{1} \;\;,\;\; \Lambda_2 = \lambda^{\prime}_{2} + \hat{\lambda}_{2}  \;\;,\;\; \Lambda_3 = \lambda_{1} + \tilde{\lambda}_{1} \;\;,\;\;
    \Lambda_4 = \lambda_{21} + \lambda_{12} + \tilde{\lambda}_{21} + \tilde{\lambda}_{12}  \;\;,\;\;\Lambda_5 = \lambda_{2} + \tilde{\lambda}_{2} \, .
\label{eq:Lambdadef}
\end{equation}
Consider now the last of Eqs.~\eqref{eq:last_two_triplet} which, for non-zero values of the VEVs $v$ and $u$,
allows for non-zero independent values for the phases $\theta_1$ and $\theta_2$. For this to happen, notice how crucial is the presence of {\em both} soft-breaking terms $\mu_1$ and $\mu_2$ in the scalar potential. If either of these parameters vanishes, we fall into a situation analogous to that of the STM where SCPV cannot occur. It is also important to stress that, from what we have just shown, SCPV can be realised in the 2STM with a softly-broken U(1) symmetry  -- this is not possible in the 2HDM~\cite{Ferreira:2010hy}, but it can already be achieved in a 3HDM for which only one of the doublets is charged under the global U(1).

In the following, we will study the properties of the 2STM, especially in what concerns its scalar mass spectrum. To do so, we need to solve the minimization conditions for both CP-conserving and CP-violating vacua. For convenience, we take as input parameters the $\Phi$ and $\Delta_{1,2}$ VEVs and solve Eqs.~\eqref{eq:last_two_triplet} with respect to mass parameters of the scalar potential. In particular, in the case of CP-conserving vacua ($\theta_1=\theta_2=0$), we have
\begin{equation}
\begin{aligned}
        m^2 &= -\lambda_0 v^2 - \dfrac{1}{2}(\Lambda_1 c_\beta^2 + \Lambda_2 s_\beta^2)u^2 + \sqrt{2}u \left( \mu_1 c_\beta + \mu_2 s_\beta \right) \, , \\
        M_{11}^2 &= -\dfrac{1}{2}\Lambda_1 v^2 - \dfrac{1}{2}( 2\Lambda_3 c_\beta^2 + \Lambda_4 s_\beta^2)u^2  + \dfrac{\sqrt{2}}{2} \dfrac{\mu_1 v^2}{u} \dfrac{1}{c_\beta} -  M_{12}^2 t_\beta  \, , \\
        M_{22}^2 &= -\dfrac{1}{2}\Lambda_2 v^2 - \dfrac{1}{2}( \Lambda_4 c_\beta^2 + 2 \Lambda_5 s_\beta^2)u^2 + \dfrac{\sqrt{2}}{2} \dfrac{\mu_2 v^2}{u} \dfrac{1}{s_\beta} -  M_{12}^2 \dfrac{1}{t_\beta}  \, ,
\end{aligned}
\label{eq:CPCcond}
\end{equation}
where the compact notation $t_\beta \equiv \tan\beta$ has been used. Notice that the soft-breaking parameters $\mu_{1,2}$ and $M_{12}^2$ will be later considered to vary together with $u$, $\tan\beta$ and the quartic couplings of $V_{\rm U(1)}$. Instead, for the CP-violating case with ($\theta_{1,2}\neq 0$) we get
\begin{equation}
\begin{aligned}
        m^2 &= -\lambda_0 v^2 - \dfrac{1}{2}(\Lambda_1 c_\beta^2 + \Lambda_2 s_\beta^2)u^2 - \sqrt{2}u \mu_1 c_\beta \dfrac{s_{\theta_1-\theta_2}}{s_{\theta_2}} \, , \\
        M_{11}^2 &= -\dfrac{1}{2}\Lambda_1 v^2 - \dfrac{1}{2}( 2\Lambda_3 c_\beta^2 + \Lambda_4 s_\beta^2)u^2 - \dfrac{\sqrt{2}}{2} \dfrac{\mu_1 v^2}{u} \dfrac{1}{c_\beta} \dfrac{s_{\theta_2}}{s_{\theta_1-\theta_2}} \, , \\
        M_{22}^2 &= -\dfrac{1}{2}\Lambda_2 v^2 - \dfrac{1}{2}( \Lambda_4 c_\beta^2 + 2 \Lambda_5 s_\beta^2)u^2 - \dfrac{\sqrt{2}}{2} \dfrac{\mu_1 v^2}{u} \dfrac{1}{t_\beta s_\beta} \dfrac{s_{\theta_1}^2}{s_{\theta_1-\theta_2}s_{\theta_2}} \, , \\
        M_{12}^2 &= \dfrac{\mu_1 v^2}{\sqrt{2} u} \dfrac{1}{s_\beta} \dfrac{s_{\theta_1}}{s_{\theta_1-\theta_2}} \;\;,\;\;
        \mu_2 = -\mu_1 \dfrac{1}{t_\beta} \dfrac{s_{\theta_1}}{s_{\theta_2}} \, .
\end{aligned}
\label{eq:CPVcond}
\end{equation}
The last two equations, which stem from the $\theta_{1,2}$ extrema conditions, determine $M_{12}^2$ and $\mu_2$ as functions of the VEVs, the soft-breaking parameter $\mu_1$ and the $\theta_{1,2}$ phases. It is worth commenting on the cases with $\theta_1=0$ or $\theta_2=0$ or $\theta_1=\theta_2$ which, according to the above equations, seem to lead to inconsistent results. These unphysical outcomes originate from the fact that the minimisation conditions \eqref{eq:CPVcond} are no longer valid in these specific cases, as they have been deduced assuming all phases generic and non-zero. For instance,  considering the $\theta_1=\theta_2=\theta$ limit from the beginning, Eqs.~\eqref{eq:CPVcond} are replaced by Eqs.~\eqref{eq:CPCcond} together with $\mu_1=\mu_2=0$, where $M_{12}^2$ becomes now a free parameter. In other words, the $\theta_1 = \theta_2$ case corresponds to a
potential with a different symmetry, $\Phi\rightarrow e^{i \alpha} \Phi$, $\Delta_{1,2}\rightarrow e^{i\beta}
\Delta_{1,2}$. Without the $\mu_{1,2}$ terms, the VEVs for the triplets will break this symmetry and yield
a massless scalar (other than the desired gauge Goldstone boson).

Let us now look at the fermion sector of the model. As in the STM case, the quark
fields only couple to the doublet $\Phi$ in an identical manner to the SM. The leptons also couple to $\Phi$ as
in the SM, and with the two triplets $\Delta_{1,2}$ similarly to Eq.~\eqref{eq:Lyuk1}. Namely, the Yukawa interactions are now
\begin{equation}
 \mathcal{L}_{\rm Yuk} = \Y_{\alpha\beta}^{\Delta_1} \,L_\alpha^T \mathcal{C}^\dag  i \tau_2 \Delta_1 \,L_\beta+\Y_{\alpha\beta}^{\Delta_2} \,L_\alpha^T \mathcal{C}^\dag  i \tau_2 \Delta_2 \,L_\beta
 {\rm H.c.}\,,\; (\alpha,\beta)=e,\mu,\tau\,.
\label{eq:Lyuk2}
\end{equation}
Notice that now the existence of two Yukawa matrices, $\Y^{\Delta_1}$ and $\Y^{\Delta_2}$, opens up the possibility of inducing CP violation in the leptonic sector from SCPV with $\theta_1 \neq \theta_2 \neq 0$, in which case the effective neutrino mass matrix becomes
\begin{align}
    m_\nu=\frac{1}{2}\left(\Y^{\Delta_1}u_1 e^{i\theta_1}+\Y^{\Delta_2}u_2 e^{i\theta_2}\right)\,.
\end{align}
Demanding CP conservation at the Lagrangian level, implies the Yukawa matrices to be real and leptonic CP violation may only arise from the VEV complex phases $\theta_{1,2}$ --  it is clear that only the phase difference $\theta_1-\theta_2$ is relevant since the lepton doublets can always be rephased as $L_\alpha \rightarrow e^{-i \theta_2/2}L_\alpha$. Thus, not only both $\theta_{1,2}$ must be non-zero, but also $\theta_1-\theta_2 \neq n\pi$, if leptonic CP violation is to be originated from the vacuum. Further, the two $\Y^{\Delta}$ matrices must be linearly independent, otherwise the complex phase would be eliminated when one rotates to the mass basis. With generic $\Y^{\Delta}$ matrices, though, the complex phase $\theta_1-\theta_2$ will irreducibly appear in the lepton mixing matrix. The situation is analogous to obtaining a complex CKM matrix in the 2HDM from SCPV -- it cannot occur if the model is endowed with $Z_2$ or global U(1) symmetries, which prevent tree-level flavour changing neutral currents by forcing each doublet to couple exclusively to fermions with the same electric charge. It is only when such symmetries are removed, and there are two distinct real Yukawa matrices for the interactions between each doublet and up and down quarks, that CP violation in the CKM matrix is obtained from SCPV. Finally, concerning the neutrino mass matrix, without any further assumptions or symmetry principle constraining the flavour structure of $\Y^{\Delta_{1,2}}$, the relation between the Dirac and Majorana CP phases appearing in the lepton mixing matrix and the VEV phases cannot be directly established. This may not be the case when flavour symmetries are considered to shape the Yukawa couplings among lepton doublets and scalar triplets.

\section{Scalar mass spectrum of the 2STM}
\label{sec:mass}

Extensions of the SM with scalar triplets boast a rich phenomenology due to the existence of new neutral, charged and doubly-charged spin-0 particles. Actually, the latter are not present when only scalar-doublet fields are added to the SM and thus the presence of scalar triplets opens up the window to interesting collider signatures~\cite{Aaboud:2018qcu,Sirunyan:2017ret}.
As already mentioned, viable EWSB requires $\sqrt{v^2+2u^2}\simeq246 \, \text{GeV}$. Furthermore, constraints on the value of the $\rho$ parameter impose $u \lesssim 8 \, \text{GeV}$. Since $u\ll v$, we find it convenient to introduce a small parameter $\varepsilon$ defined as:
\begin{equation}
    \varepsilon \equiv \frac{u}{v} \lesssim 3.3 \times 10^{-2} \, .
    \label{eq:epsdef}
\end{equation}
In the 2STM there are six neutral, three charged and two doubly-charged scalar mass eigenstates, of which one neutral and one charged are massless, corresponding to the SM Goldstone bosons. One of the physical neutral scalars is expected to be the 125~GeV Higgs boson discovered at LHC. Taking as examples the 2HDM and the one-triplet model, one expects that some of the new scalars (or even all) can be made to decouple from the SM electroweak sector. In this section, we aim at studying the general properties of the neutral and charged scalar mass spectra for both CP-conserving and CP-violating vacuum configurations of the 2STM. We are especially interested in verifying if the model exhibits decoupling features or whether, on the other hand, there are inevitably light scalars which might have escaped experimental detection.

\subsection{CP-conserving minima}

Let us first address the neutral scalar spectrum for the case with real VEVs, i.e. with $\theta_{1,2}=0$ in
Eq.~\eqref{eq:cpviolating}. Using the notation from Eqs.~\eqref{eq:fields} and~\eqref{eq:fields12}, we
consider that the neutral complex fields are decomposed as
\begin{equation}
\phi^0= \dfrac{v + \rho_0+i\eta_0}{\sqrt{2}}\;,\;
\delta^0_{1,2}=\dfrac{u_{1,2} + \rho_{1,2}+i\eta_{1,2}}{\sqrt{2}}\,,
\end{equation}
with $\rho_i$ and $\eta_i$ real. As usual, the neutral scalar mass matrix, $\mathcal{M}_0^2$, is obtained by computing the second derivatives of the scalar potential with respect to the $\rho_i$ and $\eta_i$ degrees of freedom. The resulting $6\times 6$ matrix
$\mathcal{M}_0^2$ in the basis $S=(\rho_0,\rho_1,\rho_2,\eta_0,\eta_1,\eta_2)^T$ is such that the corresponding mass term in the Lagrangian is in the form $S^T \mathcal{M}_0^2 S$. The six mass eigenstates will be denoted by $h_i^0$ with $i=1,...,6$. In terms of the parameter $\varepsilon$ defined in Eq.~\eqref{eq:epsdef}, $\mathcal{M}_0^2$ can be decomposed as:
\begin{equation}
  \mathcal{M}_0^2= (\mathcal{H}_{0} + \mathcal{P})\,v^2 \;,\; \mathcal{P}= \varepsilon \mathcal{V}_1 + \varepsilon^2 \mathcal{V}_2\;,\;  \mathcal{H}_{0} = \left( \begin{matrix}\mathcal{A} & 0\\ 0 & \mathcal{B}\end{matrix} \right)\,,
  \label{eq:M0def}
\end{equation}
where $\mathcal{H}_{0}$, $\mathcal{P}$ and $\mathcal{V}_{1,2}$ ($\mathcal{A}$ and $\mathcal{B}$) are dimensionless $6\times 6$ ($3\times 3$) real symmetric matrices. It is worth stressing that the above expressions are exact in the sense that no perturbative expansion in the small parameter $\varepsilon$ has been done. Taking into account
the minimization conditions for a CP-conserving minimum in Eqs.~\eqref{eq:CPCcond}, we find
\begin{equation}
\begin{aligned}
\mathcal{B} =
        \begin{pmatrix}
         \dfrac{2\sqrt{2}u(\mu_1c_{\beta}+\mu_2s_{\beta})}{v^2} &\!\!\! \dfrac{-\sqrt{2}\mu_1}{v} & \dfrac{-\sqrt{2}\mu_2}{v}\\[5pt]
        \dfrac{-\sqrt{2}\mu_1}{v} &\!\!\! \quad  \dfrac{\mu_1}{\sqrt{2}u c_\beta} - \dfrac{M_{12}^2 t_{\beta}}{v^2} & \dfrac{M_{12}^2}{v^2}\\[5pt]
       \dfrac{-\sqrt{2}\mu_2}{v}  & \dfrac{M_{12}^2}{v^2} & \dfrac{\mu_2}{\sqrt{2}u s_\beta} - \dfrac{M_{12}^2}{v^2 t_\beta}
       \end{pmatrix} \, ,
 \label{eq:ABcpc}
 \end{aligned}
\end{equation}
with $\mathcal{A}$ being the same as $\mathcal{B}$, except for $\mathcal{A}_{11}=2\lambda_0$. As for $\mathcal{V}_{1,2}$, they are given by
\begin{equation}
        \mathcal{V}_{1,2} = \left( \begin{matrix}\mathcal{C}_{1,2} & 0\\ 0 & 0 \end{matrix} \right)\; ,\;
%
%
\mathcal{C}_1 =
        \begin{pmatrix}
         0 & \Lambda_1 c_{\beta} & \Lambda_2 s_{\beta}  \\[6pt]
         \Lambda_1 c_{\beta}  & 0 & 0 \\[6pt]
         \Lambda_2 s_{\beta}  & 0 & 0
        \end{pmatrix} \; , \;
        \mathcal{C}_2 =
        \begin{pmatrix}
         0 & 0 & 0  \\[6pt]
         0  & 2\Lambda_3c^2_{\beta} & \Lambda_4 \, c_{\beta}s_{\beta} \\[6pt]
         0  & \Lambda_4 \, c_{\beta}s_{\beta} & 2\Lambda_5 s^2_{\beta}
        \end{pmatrix} \,,
\end{equation}
where the dimensionless parameters $\Lambda_i$ are those defined in Eqs.~\eqref{eq:Lambdadef}. As expected, $\mathcal{M}_0^2$ is block-diagonal in the chosen basis due to absence of CP violation -- the neutral imaginary components of the fields do not mix with the real ones, a hallmark of CP conservation when one is in a basis where all parameters in the potential are real, as is our case.
Direct computation shows that the full matrix $\mathcal{M}_0^2$ has one vanishing eigenvalue corresponding to the SM massless neutral Goldstone boson. This zero eigenvalue is also the only one of $\mathcal{H}_0$, hinting at the fact that, since $\varepsilon$ is a small parameter, the non-zero masses should also mainly come from the eigenvalues of $\mathcal{H}_0$. As we shall see in the next section, this turns out to not be the case when the triplet VEVs are complex. Full analytical expressions for the five non vanishing masses $m^2_{h^0_{2-6}}$ cannot be easily obtained in the most general case. However, one can adopt a simple, yet insightful, approach based on matrix-theory results and theorems which will allow us to infer certain features of the mass spectrum. Moreover, we will resort to benchmark examples to anticipate results which will be confirmed by a full general numerical treatment in Section~\ref{sec:num}.

From now on, the eigenvalues of a generic $n \times n$ matrix $\mathcal{M}$ are dubbed $\lambda_{k}(\mathcal{M})$, with $k$ running from $1$ to $n$. For the neutral mass matrix $\mathcal{M}_0^2$ the lowest eigenvalue is always identified by $k=1$ (it corresponds to the neutral Goldstone boson) -- thus we have $\lambda_1(\mathcal{H}_0+\varepsilon \mathcal{V}_1+\varepsilon^2 \mathcal{V}_2)=0$, and the remaining eigenvalues are all positive, since we are assuming to be at a minimum. Therefore the full matrix spectrum follows:
\begin{equation}
    0 <  \lambda_2(\mathcal{H}_0+\varepsilon \mathcal{V}_1
+\varepsilon^2 \mathcal{V}_2) \leq ... \leq \lambda_6(\mathcal{H}_0+\varepsilon \mathcal{V}_1
+\varepsilon^2 \mathcal{V}_2)    \, .
\end{equation}

Let us then enunciate Weyl's inequalities (see e.g. Ref.~\cite{matrix:analysis}), a theorem of linear algebra about changes in the eigenvalues of Hermitian matrices which are ``perturbed" by addition of a matrix with typical entries much smaller than the original
one.
\\
\noindent\rule{\textwidth}{0.4pt}
\noindent \textbf{Weyl's inequalities}: Let $ A, B \in \mathbb{R}^{n \times n} $ be Hermitian matrices. Each matrix can be diagonalised using a sequence $\lambda_1(A,B) \leq ... \leq \lambda_n(A,B)$ of $n$ real eigenvalues, together with an orthonormal basis of complex eigenvectors. Then, Weyl's inequalities state that
\begin{equation}
    \lambda_{n-i-j+2}(A+B) \leq \lambda_{n-i+1}(A) + \lambda_{n-j+1}(B) \, .
\end{equation}
valid whenever $i,j \geq 1$ and $i+j-1 \leq n$. \\
\hrule height 0.4pt depth 0pt width \textwidth \relax
\vspace{0.4cm}

\noindent The above result implies that if $B$ is small compared to $A$ (in the sense that all its eigenvalues are small compared to $A$'s), then the eigenvalues of $A + B$ are close to those of $A$. This is actually the case we are dealing with since the entries of $\mathcal{P}$ are, at most, of order $\Lambda_i \varepsilon \ll 1$, while the potentially large ratios $\mu_i/u$, $\mu_i/v$ and $M_{1,2}^2/v^2$ will render the eigenvalues of $\mathcal{H}_0$ large. Considering the case with $n-i-j+2=n-i+1 \equiv k$, Weyl's inequalities imply
%
$\lambda_k(A+B)-\lambda_k(A) \leq \lambda_n(B)$,
%
meaning that the difference between the $k$-th eigenvalue of $A+B$ and the $k$-th eigenvalue of $A$ is bounded by the spectral radius of $B$ (the largest absolute value of its eigenvalues).

Identifying $A$ and $B$ with $\mathcal{H}_0$ and $\mathcal{P}$ given in Eqs.~\eqref{eq:M0def}, respectively, we get
\begin{equation}
    \lambda_k(\mathcal{M}_0^2/v^2)-\lambda_k(\mathcal{H}_0) \leq \lambda_6( \mathcal{P}) \,,
    \label{eq:eigenH0weyl}
\end{equation}
where we have used the fact that the largest eigenvalue of $\mathcal{P}$ is $\lambda_6(\mathcal{P})$. As for $\mathcal{M}_0/v^2$, the matrix $\mathcal{H}_0$ has a single vanishing eigenvalue $\lambda_1(\mathcal{H}_0)=0$,
which, as we already mentioned, corresponds to the expected neutral Goldstone boson. If, in general, the magnitude of $\lambda_k \left( \mathcal{H}_0 \right)$ for $k > 1$ is much larger than that of $\lambda_6(\mathcal{P})$, then, according to Eq.~(\ref{eq:eigenH0weyl}), the difference between the eigenvalues of $\mathcal{M}_0/v^2$ and the ones of $\mathcal{H}_0$ will be relatively small. Thus, we expect the non-vanishing scalar masses to stem mainly from the five non-vanishing eigenvalues of $\mathcal{H}_0$, as anticipated above.

Under this premise, the neutral scalar masses are approximately given by
\begin{equation}
\begin{split}
      m_{h_2^0}^2 \simeq & \, 2 \lambda_0 v^2 - 2 \sqrt{2} u \bar{\mu} \, ,\quad \bar{\mu}=\mu_1 c_\beta + \mu_2 s_\beta\\
      m_{h_{3,4}^0}^2 \simeq & \, - \frac{M_{12}^2}{s_{2\beta}} + \frac{\sqrt{2}}{2} \frac{v^2 \bar{\mu}}{u s_{2\beta}}
      +  \,  \sqrt{\frac{v^2 }{u^2s_{2\beta}} \left( -\mu_1 \mu_2 v^2 + \sqrt{2} M_{12}^2 u \bar{\mu} \right) + \left( - \frac{M_{12}^2}{s_{2\beta}} + \frac{\sqrt{2}}{2} \frac{v^2 \bar{\mu}}{u s_{2\beta}} \right)^2 }\\
      m_{h_{5,6}^0}^2 \simeq & \, - \frac{M_{12}^2}{s_{2\beta}} + \frac{\sqrt{2}}{2} \frac{v^2 \bar{\mu}}{u s_{2\beta}}
      -  \,  \sqrt{\frac{v^2 }{u^2s_{2\beta}} \left( -\mu_1 \mu_2 v^2 + \sqrt{2} M_{12}^2 u \bar{\mu} \right) + \left( - \frac{M_{12}^2}{s_{2\beta}} + \frac{\sqrt{2}}{2} \frac{v^2 \bar{\mu}}{u s_{2\beta}} \right)^2 }\,.
\end{split}
\label{eq:CPc_exact}
\end{equation}
The above equations show one can choose a region of parameter space that yields a decoupling limit. To wit, since $|u| \ll v$, taking $u \bar{\mu}$ to be, at most, of order $\mathcal{O}(v^2)$ and $|M_{12}^2| \gg v^2$, we see that we obtain one neutral scalar with mass $m_{h_2^0}\sim v$, and two pairs of quasi-degenerate states whose masses can be arbitrarily large, since both $M^2_{12}$ and $\bar{\mu}/u$ can have arbitrarily large values. Aditionally, we can show that the states $h^0_3$ and $h^0_5$ are CP-even, whereas $h^0_4$ and $h^0_6$ are CP-odd and, thus,
pseudoscalars.

We now turn our attention to the three singly- and two doubly-charged scalars of the 2STM, labelled as $H_i^+$ $(i=1,2,3)$ and $H_i^{++}$ $(j=1,2)$, respectively. The $H_i^+$ masses are determined by the $3 \times 3$ matrix
$\mathcal{M}_{+}^2$ which, similarly to $ \mathcal{M}_{0}^2$, can be decomposed as
\begin{equation}
 \mathcal{M}_{+}^2 = \left(\mathcal{H}_0^+ + \mathcal{H^+_\lambda} + \varepsilon \mathcal{V}_1^+ + \varepsilon^2 \mathcal{V}_2^+\right)v^2\,,
\end{equation}
where $\varepsilon$ is, once more, the small parameter defined in Eq.~\eqref{eq:epsdef}. The zeroth order matrices in $\varepsilon$ are given by
\begin{equation}
    \mathcal{H}_0^+ =
    \begin{pmatrix}
        \dfrac{\sqrt{2}u (\mu_1 c_\beta + \mu_2 s_\beta)}{v^2} & -\dfrac{\mu_1}{v} & - \dfrac{\mu_2}{v} \\[10pt]
        -\dfrac{\mu_1}{v} & \dfrac{\mu_1}{\sqrt{2}u c_\beta} - \dfrac{M_{12}^2 t_\beta}{v^2} & \dfrac{M_{12}^2}{v^2} \\[10pt]
        -\dfrac{\mu_2}{v} & \dfrac{M_{12}^2}{v^2} & \dfrac{\mu_2}{\sqrt{2}u s_\beta}-\dfrac{M_{12}^2}{v^2 t_\beta}
    \end{pmatrix} \, , \, \mathcal{H}_\lambda^+={\rm diag}\left(0,- \dfrac{\hat{\lambda}_{1}}{4},- \dfrac{\hat{\lambda}_{2}}{4} \right)
\end{equation}
while $\mathcal{V}_{1,2}^+$ are written as
\begin{equation}
       \mathcal{V}_1^+ =
    \begin{pmatrix}
     0 & \dfrac{\hat{\lambda}_{1} c_\beta}{2\sqrt{2}} & \dfrac{\hat{\lambda}_{2} s_\beta}{2\sqrt{2}} \\[7pt]
     \dfrac{\hat{\lambda}_{1} c_\beta}{2\sqrt{2}} & 0 & 0 \\[7pt]
     \dfrac{\hat{\lambda}_{2} s_\beta}{2\sqrt{2}} & 0 & 0
     \end{pmatrix} \, , \,
     \mathcal{V}_2^+ =
     \begin{pmatrix}
     -\dfrac{\hat{\lambda}_{1}c_\beta^2 + \hat{\lambda}_{2}s_\beta^2}{2} & 0 & 0 \\[7pt]
     0 & - \dfrac{\Lambda_6 s_\beta^2}{4} & \dfrac{\Lambda_6 s_{2\beta}}{8}\\[7pt]
     0 & \dfrac{\Lambda_6  s_{2\beta}}{8} & -\dfrac{\Lambda_6  c_\beta^2}{4}
     \end{pmatrix} \, ,
\end{equation}
with $ \Lambda_6=2 \lambda_{12} + \tilde{\lambda}_{21} + \tilde{\lambda}_{12} $.

The full matrix $\mathcal{M}_{+}^2$ has a vanishing eigenvalue, which is identified with the charged SM
Goldstone boson, and $\mathcal{H}_0^+$ also has a single vanishing eigenvalue. Thus, using the same arguments
as before, the singly-charged spectrum can be approximated by the spectrum of the matrix $\mathcal{H}_0^+ + \mathcal{H}_\lambda^+$,
since the contributions from  $\mathcal{V}_1^+$ and $\mathcal{V}_2^+$ are affected by powers of $\varepsilon$. In fact,
we can further argue that the singly-charged masses are, to very good approximation, given by the eigenvalues
of $\mathcal{H}_0^+$,
since the non-zero elements of $\mathcal{H}_\lambda^+$ are much smaller in magnitude than the $(2,2)$ and $(3,3)$
entries of $\mathcal{H}_0^+$ when $|M^2_{12}|\gg v^2$,  which is in the decoupling region we found for the neutral
scalars~\footnote{This can also be understood as another application of Weyl's inequalities, which
we leave as an exercise for the reader.}. Thus, we obtain the following very good approximations for
the charged masses,
\begin{equation}
   m_{H_1^+}^2 = 0 \quad , \quad m_{H_2^+}^2 \approx m_{h_{3}^0}^2 \quad , \quad m_{H_3^+}^2 \approx m_{h_{5}^0}^2 \,,
\end{equation}
where the vanishing eigenvalue corresponds, as we already mentioned, to the expected Goldstone boson. Notice the degeneracy between neutral and charged particles, where $m_{h_{3,5}^0}^2$ are approximately given as shown in Eqs.~(\ref{eq:CPc_exact}). A full numerical calculation shows that this degeneracy is not exact -- it is lifted by terms of order $\mathcal{O}(\varepsilon)$ -- but it is an excellent approximation nonetheless.

To conclude the CP-conserving mass spectrum analysis, we now address the doubly-charged spectrum. There are two doubly-charged mass states, $H_i^{++}$, with $i=1,2$, which stem from the triplet fields' doubly-charged fields,
$\delta_i^{++}$ as in Eq.~\eqref{eq:fields12}. Their squared mass matrix can be written as
\begin{equation}
   \frac{\mathcal{M}_{++}^2}{v^2} = \mathcal{H}_0^{++} + \mathcal{H}_\lambda^{++} + \varepsilon^2 \mathcal{V}_2^{++} \, ,
\end{equation}
and these submatrices are given by
\begin{equation}
    \mathcal{H}_0^{++} =
    \begin{pmatrix}
        \dfrac{\mu_1}{\sqrt{2}u c_\beta} - \dfrac{M_{12}^2 t_\beta}{v^2} & \dfrac{M_{12}^2}{v^2} \\[10pt]
        \dfrac{M_{12}^2}{v^2} & \dfrac{\mu_2}{\sqrt{2}u s_\beta}-\dfrac{M_{12}^2}{v^2 t_\beta}
    \end{pmatrix} \, ,
\end{equation}
\begin{equation}
 \mathcal{H}_\lambda^{++} =
    \begin{pmatrix}
        - \dfrac{\hat{\lambda}_{1}}{2} & 0 \\[10pt]
        0 & - \dfrac{\hat{\lambda}_{2}}{2}
    \end{pmatrix} \, , \,
    \mathcal{V}_2^{++} =
    \begin{pmatrix}
        - \tilde{\lambda}_{1}c_\beta^2 - \dfrac{\Lambda_7}{2} s_\beta^2 & \dfrac{\lambda_{12}}{2} c_\beta s_\beta \\[10pt]
        \dfrac{\lambda_{12}}{2} c_\beta s_\beta & - \tilde{\lambda}_{2} s_\beta^2 - \dfrac{\Lambda_7}{2} c_\beta^2
    \end{pmatrix} \, ,
\end{equation}
with $\Lambda_7=\lambda_{12}+\tilde{\lambda}_{21}+\tilde{\lambda}_{12}$.

The matrix $\mathcal{H}_0^{++}$ does not have a vanishing eigenvalue as was to be expected. Following the analysis of the singly-charged mass spectrum, we can approximate the eigenvalues of the full matrix $\mathcal{M}_{++}^2$ by the ones stemming
from just $\mathcal{H}_0^{++}$. This way, we are able to obtain approximate (albeit very accurate) expressions for
the doubly-charged  squared masses, from which the following quasi degeneracies stand out:
\begin{equation}
    m_{h^0_{3,4}}^2  \approx m_{H_2^+}^2 \approx m_{H_1^{++}}^2 \quad , \quad  m_{h^0_{5,6}}^2  \approx m_{H_3^+}^2 \approx m_{H_2^{++}}^2 \, .
    \label{eq:cpcdeg}
\end{equation}
Therefore, in the region where $|M^2_{12}|\gg v^2$, the CP-conserving mass spectrum has one scalar with mass of $\mathcal{O}(v)$, while the remaining states, both neutral or charged, can be made much heavier and, thus, decouple from the theory. We observe that we obtain two sets of quasi-degenerate particles, each set containing a CP-even scalar ($h^0_3$ or $h^0_5$), a pseudoscalar ($h^0_4$ or $h^0_6$), a singly-charged scalar ($H_2^+$ or $H_3^+$) and a doubly charged one ($H_1^{++}$ or $H_2^{++}$). As further confirmation, we numerically checked that the approximated expressions for the masses stemming from $\mathcal{H}_0$-like matrices hold in the region of parameter space considered and hence the spectrum exhibits the features described above.

\subsection{CP-violating minima}

We now draw our attention to the case of complex VEVs defined by Eqs.~(\ref{eq:cpviolating}), to which correspond the five independent minimisation conditions of Eqs.~(\ref{eq:last_two_triplet}). It is worth reminding that, in contrast with the CP-conserving case, the soft-breaking parameters $M_{12}^2$ and $\mu_2$ can now be expressed as function of the remaining parameters -- see Eqs.~\eqref{eq:CPVcond}. Once more, we write the neutral scalar mass matrix in the form of Eq.~\eqref{eq:M0def} where now:
\begin{equation}
    \mathcal{A},\mathcal{B} =
        \begin{pmatrix}
         2 \lambda_0 \, , \,  -\dfrac{2\sqrt{2}\mu_1u c_\beta s_{\theta_1-\theta_2}}{v^2 s_{\theta_2}}& \dfrac{-\sqrt{2}\mu_1}{v} & \dfrac{\sqrt{2}\mu_1 s_{\theta_1}}{v t_\beta s_{\theta_2}}\\[10pt]
        \dfrac{-\sqrt{2}\mu_1}{v} &  -\dfrac{\mu_1 s_{\theta_2}}{\sqrt{2}u s_{\theta_1-\theta_2} c_{\beta}} & \dfrac{\mu_1 s_{\theta_1}}{\sqrt{2}u s_{\beta} s_{\theta_1-\theta_2}}\\[10pt]
        \dfrac{\sqrt{2}\mu_1 s_{\theta_1}}{v t_\beta s_{\theta_2}}  & \dfrac{\mu_1 s_{\theta_1}}{\sqrt{2}u s_{\beta} s_{\theta_1-\theta_2}} & -\dfrac{\mu_1    s^2_{\theta_1}}{\sqrt{2}u t_{\beta} s_{\beta} s_{\theta_1-\theta_2} s_{\theta_2} }
        \end{pmatrix} \, .
        \label{eq:ABcpv}
\end{equation}
Due to the fact that $\theta_{1,2}\neq 0$, the matrices $\mathcal{V}_{1,2}$ are no longer block diagonal as for the CP-conserving case -- this is a clear indication of CP violation, since the real and imaginary neutral components of
the doublet and triplets have opposite CP quantum numbers. A simple calculation shows that $\mathcal{V}_1$ is
such that
\begin{equation}
(\mathcal{V}_1)_{1j}=  (\mathcal{V}_1)_{j1} \equiv V_j=(0,\Lambda_1 c_{\beta} c_{\theta_1},\Lambda_2 s_{\beta} c_{\theta_2}, 0, \Lambda_1 c_{\beta} s_{\theta_1}, \Lambda_2 s_{\beta} s_{\theta_2} )_j\,,
 \label{eq:V1cpv}
\end{equation}
%
%
%
and all remaining entries are zero (notice the vanishing elements of the fourth line and column of $\mathcal{V}_1$). As for $\mathcal{V}_2$ we have:
\begin{equation}
    \mathcal{V}_2 =
        \begin{pmatrix}
         0 & 0 & 0 & 0 & 0 & 0 \\[6pt]
         0 & 2\Lambda_3 c^2_{\beta}c^2_{\theta_1} & \Lambda_4 \, c_{\beta}c_{\theta_1}c_{\theta_2}s_{\beta} & 0 &  \Lambda_3c^2_{\beta}s_{2\theta_1} & \Lambda_4 \, c_{\beta}c_{\theta_1}s_{\theta_2}s_{\beta}\\[6pt]
         0 & \Lambda_4 \, c_{\beta}c_{\theta_1}c_{\theta_2}s_{\beta} & 2\Lambda_5 s^2_{\beta}c^2_{\theta_2} & 0 & \Lambda_4 \, c_{\beta}c_{\theta_2}s_{\theta_1}s_{\beta} & \Lambda_5 s^2_{\beta}s_{2\theta_2}\\[6pt]
         0 & 0 & 0 & 0 & 0 & 0\\[6pt]
         0 & \Lambda_3 \Lambda_3 c^2_{\beta}s_{2\theta_1} & \Lambda_4 \, c_{\beta}c_{\theta_2}s_{\theta_1}s_{\beta} & 0 & 2 \Lambda_3 c^2_{\beta}s^2_{\theta_1} & \Lambda_5 \, c_{\beta}s_{\theta_1}s_{\theta_2}s_{\beta}\\[6pt]
         0 & \Lambda_4 \, c_{\beta}c_{\theta_1}s_{\theta_2}s_{\beta} & \Lambda_5 s^2_{\beta}s_{2\theta_2} & 0 & \Lambda_5 \, c_{\beta}s_{\theta_1}s_{\theta_2}s_{\beta} & 2 \Lambda_5 s^2_{\beta}s^2_{\theta_2}
        \end{pmatrix} \, .
         \label{eq:V2cpv}
\end{equation}

In the present scenario $\mathcal{H}_0$ has three vanishing eigenvalues, $\lambda_{1,2,3}(\mathcal{H}_0)=0$, instead of just one as in the real-vacua case. Resorting once more to Weyl's inequalities in the form of Eq.~(\ref{eq:eigenH0weyl}), and taking into account that $\lambda_{1}(\mathcal{M}_0^2)=0$ (corresponding to the SM neutral Goldstone boson), we
can write
%
\begin{equation}
    \lambda_k(\mathcal{M}_0^2/v^2) \leq \lambda_6(\mathcal{P}) \, ,k=2,3.
\end{equation}
Thus, we can already anticipate that $\mathcal{P}= \varepsilon \mathcal{V}_1 + \varepsilon^2 \mathcal{V}_2$ shall be responsible for, somehow, giving rise to two light neutral scalars and, since the matrix $\mathcal{P}$ is of order $\sim \varepsilon$, the three lightest eigenvalues of the full mass matrix $\mathcal{M}_0^2$ will be bounded from above by masses which we expect to be small when compared with the electroweak scale. We therefore see the possibility of a non-decoupling regime emerging. It turns out that, as we will now show, the full neutral spectrum includes two light scalars expected to have a mass upper bound coming just from the $\varepsilon^2 \mathcal{V}_2$ matrix. Following the previous reasoning, the remaining three mass states (in which the 125~GeV Higgs boson is included) stem mainly from the non-vanishing eigenvalues of $\mathcal{H}_0$, expected to be much larger than those of $\mathcal{P}$. To prove this, we start by considering the full mass matrix decomposed as in the CP conserving case:
\begin{equation}
\mathcal{M}_0^2= (\mathcal{H}_{0} + \varepsilon \mathcal{V}_1 + \varepsilon^2 \mathcal{V}_2)\,v^2 \;,
\end{equation}
and make use of another result from matrix theory: Cauchy's Eigenvalue Interlacing Theorem (EIT)~\cite{matrix:analysis}.

\noindent\rule{\textwidth}{0.4pt}

\noindent\textbf{Eigenvalue Interlacing Theorem (EIT):} Let $ A \in \mathbb{R}^{n \times n} $ be a symmetric matrix, and $ B_i \in \mathbb{R}^{m \times m} $ (with $m<n$) a principal submatrix of $A$ obtained by removing its $i$-th row and the $i$-th column. Denoting the eigenvalues of $A$ and $B_i$ by $\lambda_1(A) \leq ... \leq \lambda_n(A)$ and $\lambda_1(B_i) \leq ... \leq \lambda_m(B_i)$, respectively, then
\begin{align}
    \lambda_k(A) \leq \lambda_k(B_i) \leq \lambda_{k+n-m}(A) \, ,\quad k=1,...,m\,.
\end{align}
\hrule height 0.4pt depth 0pt width \textwidth \relax
\vspace{0.3cm}
Let us apply this theorem to the $6\times 6$ normalised mass matrix $\mathcal{M}_0^2/v^2$, which we identify with $A$ in the EIT. Removing the first line and column of $\mathcal{M}_0^2/v^2$ (and denoting with a tilde any matrix resulting from this operation), we define the principal $5\times 5$ submatrix $\tilde{\mathcal{M}}_0^2/v^2=\tilde{\mathcal{H}}_0+\tilde{\mathcal{P}}$. Given that $\tilde{\mathcal{V}}_1=0$ [see Eq.~\eqref{eq:V1cpv}] we get $\tilde{\mathcal{M}}_0^2/v^2=\tilde{\mathcal{H}}_{0} + \varepsilon^2 \tilde{\mathcal{V}}_2$. Considering the EIT with $n=6$ and $m=5$, we have $\lambda_k(\mathcal{M}_0^2/v^2) \leq  \lambda_k(\tilde{\mathcal{H}}_{0} + \varepsilon^2 \tilde{\mathcal{V}}_2) \leq \lambda_{k+1}(\mathcal{M}_0^2/v^2)$ which, for $k=2,3$, implies $   \lambda_{2,3}(\mathcal{M}_0^2/v^2) \leq  \lambda_{2,3}(\tilde{\mathcal{H}}_{0} + \varepsilon^2 \tilde{\mathcal{V}}_2) \leq \lambda_{3,4}(\mathcal{M}_0^2/v^2)$. Since $\lambda_k(\mathcal{M}_0^2/v^2) >0$ for $k>1$ (all squared masses are positive, since we are assuming that these VEVs correspond to a minimum), we conclude that
\begin{equation}
    0 < \lambda_{2,3}(\mathcal{M}_0^2/v^2) \leq  \lambda_{2,3}(\tilde{\mathcal{H}}_{0} + \varepsilon^2 \tilde{\mathcal{V}}_2) \, ,
\end{equation}
i.e., the second and third largest eigenvalues of the full matrix are bounded from above by $\lambda_2(\tilde{\mathcal{H}}_{0} + \varepsilon^2 \tilde{\mathcal{V}}_2)$ and $\lambda_3(\tilde{\mathcal{H}}_{0} + \varepsilon^2 \tilde{\mathcal{V}}_2)$, respectively. A simple analytical expression for such upper bounds is hard to obtain in the most general case. However, recalling Weyl's Inequalities, we can write:
\begin{equation}
    \lambda_{2,3}(\tilde{\mathcal{H}}_{0} + \varepsilon^2 \tilde{\mathcal{V}}_2) \leq
   \lambda_{2,3}(\tilde{\mathcal{H}}_{0}) + \varepsilon^2 \lambda_{4,5}(\tilde{\mathcal{V}}_2) \, ,
   \label{eq:tildelimit}
\end{equation}
where $\lambda_{4,5}(\tilde{\mathcal{V}}_2)$ stand for the largest two eigenvalues of the $5 \times 5$ matrix $\tilde{\mathcal{V}}_2$. Since $\tilde{\mathcal{H}}_{0}$ has three vanishing and two non-vanishing eigenvalues it is straightforward to conclude that $\lambda_{3}(\tilde{\mathcal{H}}_{0})=0$, regardless of the sign of both non-vanishing eigenvalues. Although one cannot guarantee that $\lambda_{2}(\tilde{\mathcal{H}}_{0})=0$ (since there can be negative eigenvalues), one can show that this is actually the case when requiring all masses to be positive. Thus, in the case of physical interest, $\lambda_{2}(\tilde{\mathcal{H}}_{0})=\lambda_{3}(\tilde{\mathcal{H}}_{0})=0$ and Eq.~\eqref{eq:tildelimit} imply an upper bound  on the lightest-scalar masses $m^2_{h_{2,3}^0}$:
\begin{equation}
   m^2_{h_{2,3}^0}= \lambda_{2,3}(\mathcal{M}_0^2) \leq u^2\lambda_{4,5}(\tilde{\mathcal{V}}_2) \,,
\end{equation}
which, after extracting $\tilde{\mathcal{V}}_2$ from $\mathcal{V}_2$ given in Eq.~\eqref{eq:V2cpv}, leads to
\begin{equation}
     m^2_{h_{2,3}^0} \leq \frac{u^2}{2} \left[
    \Lambda_3 + \Lambda_5 + (\Lambda_3-\Lambda_5)c_{2\beta}\pm\sqrt{\left[\Lambda_3 + \Lambda_5 + (\Lambda_3-\Lambda_5)c_{2\beta}\right]^2 + \left(
   \Lambda_4^2 -4\Lambda_3\Lambda_5\right)s^2_{2\beta}}\right] \, .
    \label{eq:CP_lim3}
\end{equation}
This result implies that there is an upper bound on the two lightest neutral scalar masses proportional to $f(\Lambda)u$, where $f(\Lambda)$ is a function of $\Lambda_i$ and $\beta$ defined in Eqs.~\eqref{eq:Lambdadef} and \eqref{eq:cpviolating}, respectively. Considering all quartic parameters in the scalar potential to be at most 10, we arrive at a rough estimate of $m^2_{h_2^0} \lesssim 3u$ and $m^2_{h_3^0} \lesssim 6u$. Having in mind that EWPD sets $u\leq 8$~GeV, we conclude that $m^2_{h_3^0} \lesssim 25$~GeV and $m^2_{h_3^0} \lesssim 50$~GeV. This result, obtained resorting to simple linear algebra theorems, turns out to be in excellent agreement with the numerical simulations presented in Section~\ref{sec:num}.

This analytical result also casts a different light on the possibility of generating naturally small neutrino masses in the 2STM with $\mathcal{O}(1)$ Yukawa couplings and triplet VEVs $u$ of the eV order -- we now see that, if CP is broken spontaneously, there will be two extremely light neutral scalars. In short, in a scenario where the triplet VEVs are complex and their magnitude is of the order of the neutrino mass scale, a scalar spectrum with all particles much heavier than the SM Higgs boson cannot be realised. It should be stressed that, even if $u$ is close to its 8 GeV upper bound, two of the neutral scalar masses will be at most a few tenths of GeV. This, however, would require very tiny triplet Yukawa couplings so that $\mathcal{O}$(eV) neutrino masses are generated.

From the analysis above, we conclude that the main contribution to the masses of the three heaviest neutral scalars stem from $\mathcal{H}_0$, which has the block diagonal form shown in Eq.~\eqref{eq:M0def} with $\mathcal{A}$ and $\mathcal{B}$ defined in Eq.~\eqref{eq:ABcpv}. One of those masses (typically the lightest) correspond to the 125~GeV SM Higgs mass, while the remaining two are heavier.
Barring accidental and/or special $\tan\beta$ limits, the entries of $\mathcal{A}$ and $\mathcal{B}$ scale as $\mu_1/u$ and $\mu_1/v$. Given that $u \ll v$, the former ratio is much larger than the latter. We can
obtain the following approximate expressions for the neutral masses,
\begin{equation}
\begin{aligned}
m_{h^0_4}^2 &\simeq 2\lambda_0 v^2 + 2 \sqrt{2} \, \mu_1 u c_\beta \dfrac{s_{\theta_1-\theta_2}}{s_{\theta_2}}  \,,\\
 m_{h^0_5}^2 &\simeq m_{h^0_6}^2 \simeq - \dfrac{s_{\theta_2}}{s_{\theta_1-\theta_2}} \left[ f_1(\beta,\theta_1,\theta_2) +  f_2(\beta) \right] M_\Delta^2 \,,
\end{aligned}
\label{eq:neutral_high_masses_CP}
\end{equation}
where, similarly to what we have done for the STM model, we now define $M_\Delta$ as
\begin{equation}
    M_\Delta^2 \equiv \dfrac{\mu_1\,v^2}{\sqrt{2}u}   \, .
    \label{eq:Mdelta}
\end{equation}
The functions $f_1$ and $f_2$ are given by:
\begin{equation}
    f_1(\beta, \theta_1,\theta_2)=\dfrac{s^2_{\theta_1}}{t_\beta s_\beta s^2_{\theta_2}} \quad , \quad f_2(\beta) =\dfrac{1}{c_\beta} \, .
    \label{eq:f12def}
\end{equation}
From Eq.~\eqref{eq:neutral_high_masses_CP} we see that we can obtain a SM-like scalar state
-- corresponding to $h^0_4$ --  with a mass close to the  electroweak scale provided $\mu_1 u \lesssim \mathcal{O}(v^2)$. We  will verify later that this choice is also  compatible with $h^0_4$ having couplings
to gauge bosons and fermions similar to those of the SM Higgs. Further, we see from the same equation that if we
wish to have the two extra neutral scalars be much heavier than the weak scale (and thus decouple, easily
avoiding current LHC search limits) we need to ensure $M^2_\Delta$ is large enough, which pushes us
to the regime where $\mu_1/u$ assumes large values. To quantify this statement, having $M_\Delta$ be of the TeV
order would require (modulo large enhancements/suppressions from the angular factors in
Eq.~\eqref{eq:neutral_high_masses_CP}) $\mu_1 \gtrsim 25 u$.

At this point, it is convenient to comment on the special case $\theta_1=\theta_2$ which, according to the above result, seems to lead to infinite $h_{5,6}^0$ masses. As we mentioned above, this situation corresponds to $\mu_1 = \mu_2 = 0$, which means that the U(1) symmetry is now exact, and a VEV for the doublet yields two zero eigenvalues for the neutral scalars -- one the gauge Goldstone boson, the other a pseudo-Goldstone. This is an extreme case of non-decoupling, wherein one of the extra scalars is actually massless, and we will not consider it further.

We now comment on the charged-scalar sector of the 2STM, which comprises two singly and two doubly-charged physical
states (besides the charged massless Goldstone bosons) denoted by $H_{1,2}^+$ and $H_{1,2}^{++}$, respectively. The
corresponding mass matrices are determined, as usual, by the second derivatives of the scalar potential with respect to
the $\delta_{1,2}^+$ and $\delta_{1,2}^{++}$ degrees of freedom -- see Eq.~\eqref{eq:fields12}. Considering the
case $\mu_1 \gg u$, and Taylor expanding the eigenvalues of the mass matrices in the small parameter $u/\mu_1$,
we get the following approximate masses for the charged physical scalars:
\begin{equation}
\begin{aligned}
  m^2_{H_2^+} &\simeq  \frac{m^2_{H_1^{++}}}{2} \simeq -\dfrac{1}{4} \dfrac{\hat{\lambda}_{1}f_1(\beta,\theta_1,\theta_2)+\hat{\lambda}_{2}f_2(\beta)}{f_1(\beta,\theta_1,\theta_2)+f_2(\beta)} v^2 \,,\\
  m^2_{H_3^+} &\simeq  m^2_{H_2^{++}} \simeq - \dfrac{s_{\theta_2}}{s_{\theta_1-\theta_2}} \left[ f_1(\beta,\theta_1,\theta_2)+f_2(\beta) \right] M_\Delta^2 \,,
\end{aligned}
\label{eq:mcCPV}
\end{equation}
with $f_{1,2}$ and $M_\Delta^2$ defined as in Eqs.~\eqref{eq:Mdelta} and \eqref{eq:f12def}, respectively. These results show that there is a pair of charged scalars $H_1^+$ and $H_1^{++}$ with similar masses set by the electroweak scale $v$, while $H_2^+$ and $H_2^{++}$ decouple in the limit of very large $M_{\Delta}^2$.

To conclude, in the case of CP-violating vacua, and assuming $\mu_1\gg u$ so as to have at least {\em some} decoupled
states, the 2STM scalar mass spectrum exhibits the following features:
\begin{itemize}
\item Besides the 125 GeV SM Higgs $h_4^0$, there are two light neutral scalars $h_{2,3}^0$ with masses proportional to the triplet VEV parameter $u$ meaning that these particles cannot mix significantly with SM fermions and gauge bosons. The remaining two scalars $h_{5,6}^0$ have similar masses and scale with $\mu_1 v^2/u$, which is typically very large.
\item There are four charged scalars $H_{1,2}^+$ and $H_{1,2}^{++}$ with $m^2_{H_1^{++}}\simeq 2 m^2_{H_1^+} \sim v^2$
and   $m^2_{H_2^+} \simeq  m^2_{H_2^{++}} \sim \mu_1 v^2/u$, i.e. one singly-charged and one doubly-charged scalars have
masses of the order of the electroweak scale, while the remaining two can be much heavier. In
addition, $H_2^+$ and $H_2^{++}$ have masses similar to those of $h_{5,6}^0$.
\item It is interesting to notice that, while for the CP conserving vacuum we obtained quasi degeneracy between
two pairs of singly-charged-doubly-charged scalar (see Eq.~\eqref{eq:cpcdeg}), for the CP-violating vaccum
the mass of one pair remains the same (for $H_3^+$ and $H_2^{++}$), but for the other pair (for $H_2^+$ and $H_1^{++}$)
the doubly-charged scalar mass is approximately twice of the singly-charged one.
 \end{itemize}
Thus, and unlike the CP-conserving case, there is no region of parameter space for which all extra scalars --
neutral or charged -- can be guaranteed to have masses much above the electroweak scalar, and the 2STM with SCPV
has no decoupling limit. In Table~\ref{tab:final_results} we summarise the properties of the neutral and scalar mass spectrum for both the CP-conserving and the CP-violating vacuum configurations.

Before moving to the numerical analysis which will confirm the results of this section, let us briefly comment on the relation between non-decoupling and SCPV in the 2STM, when compared to the 2HDM case. In Ref.~\cite{Nebot:2020niz} it was shown that, when SCPV occurs in the 2HDM, the extra scalar masses cannot be made arbitrarily large. Rather, there are upper bounds on those masses set by the fact that the size of the quartic couplings is limited by perturbativity and unitarity. That the 2HDM has a decoupling limit in the CP-conserving case but not in the CP-violating one is due to the different number of minimisation conditions -- two and three, respectively. Since the model has only {\em three} quadratic parameters, a CP-conserving minimum fixes two of them to be of order $\lambda_i v^2$, with $\lambda_i$ a generic quartic coupling of limited magnitude. A third quadratic coefficient of the scalar potential becomes, therefore, a free parameter and can be used to render the extra scalar masses as heavy as wanted, thus yielding a decoupling regime. However, in the case of SCPV the extra minimisation equation will force all quadratic terms in the potential to be of order $\lambda_i v^2$, which becomes also the mass scale of all (quadratic) scalar masses and sets an upper bound for those masses.

In the 2STM the extra minimisation equations needed for SCPV -- five equations, as in~\eqref{eq:last_two_triplet}, as opposed to three for a CP-conserving extremum, as in~\eqref{eq:CPCcond} -- do limit the size of an extra two mass parameters, compared to the CP-conserving case. But there is a difference regarding the 2HDM case: if one looks at the fully softly-broken potential of Eqs.~\eqref{eq:Vs} and~\eqref{eq:SB12}, one counts a total of {\em six} mass parameters. Even in the SCPV case, the minimisation conditions will only fix the magnitude of at most {\em five} of them, so {\em a priori} one would have a free mass parameter which could drive the extra scalar masses as high as wanted. But in the 2STM that does not occur because electroweak precision data requires that the triplet VEVs be small, and that extra requirement, along with the additional minimisation conditions, forces the model to be non-decoupling when SCPV occurs.

\setlength{\tabcolsep}{10pt}
\renewcommand{\arraystretch}{1.5}
\begin{table}
\centering
\begin{tabular}{ c|c|c c }
\multicolumn{2}{c}{Mass spectrum} & CP-Conserving & CP-Violating \\
\hline
\multirow{6}{*}{Neutral} & $h_1^0$ & \multicolumn{2}{c}{Massless - Goldstone boson} \\
& $h_2^0$ & SM Higgs-like &   \\
& $h_3^0$ & \cellcolor{green!70!orange!30!}  & \multirow{-2}{*}{Light} \\
& $h_4^0$ & \multirow{-2}{*}{\cellcolor{green!70!orange!30!}Decoupled} & SM Higgs-like\\
& $h_5^0$ & \cellcolor{orange}  & \cellcolor{cyan}  \\
& $h_6^0$ & \multirow{-2}{*}{\cellcolor{orange}Decoupled}  & \multirow{-2}{*}{\cellcolor{cyan}Decoupled}    \\
\hline
\multirow{3}{*}{Singly-charged} & $H_1^+$ & \multicolumn{2}{c}{Massless - Goldstone boson} \\
& $H_2^+$ & \cellcolor{green!70!orange!30!}Decoupled & Electroweak \\
& $H_3^+$ & \cellcolor{orange} Decoupled & \cellcolor{cyan} Decoupled\\
\hline
\multirow{2}{*}{Doubly-charged} & $H_1^{++}$ & \cellcolor{green!70!orange!30!}Decoupled & Electroweak \\
& $H_2^{++}$ & \cellcolor{orange} Decoupled & \cellcolor{cyan} Decoupled \\
\end{tabular}
\caption{Main features of the neutral and charged-scalar mass spectrum in the 2STM for the CP-conserving and CP-violating vacuum cases. Scalars marked with the same colours have masses of similar magnitude. Here,``Decoupled" is applicable to the states for which the mass can be arbitrary large, while ``Electroweak" is used for those whose mass is, typically, around the electroweak scale.}
\label{tab:final_results}
\end{table}

\section{Numerical and phenomenological  analysis}
\label{sec:num}

Having found analytical arguments for the fact that the 2STM has a non-decoupling regime when CP is spontaneously broken, we now corroborate this by performing a numerical study of the model. We also wish to investigate the possibility that the lighter scalars predicted by the analysis of the previous section might have escaped detection thus far. For that, we will concentrate on triplet VEVs of GeV order, which would require suppressed Yukawa couplings to attain small neutrino masses. Instead, with $\mathcal{O}(1)$ Yukawa couplings, $\mathcal{O}$(eV) triplet VEVs are needed, implying the existence of scalars with masses of the same order, a possibility which seems bizarre, and that deserves further investigation elsewhere. In the following, we define the mass of a generic scalar state $h$ as $m_h$. We start by considering the $\text{U}(1)$ softly-broken scalar potential of Eqs.~\eqref{eq:Vs} and~\eqref{eq:SB12}, and generate a large sample of parameters satisfying the following conditions:

\begin{itemize}
    \item The triplet VEV magnitude $u$ is chosen randomly in the interval $]0,10] \, \text{GeV}$;
    \item The VEV of the neutral component of $\Phi$ is $v = \sqrt{246^2-2 u^2}$;
    \item The angle $\beta=\arctan(u_2/u_1)$ is chosen randomly in $\beta \in \, ]0,\pi/2[$;
    \item The CP-violating phases $\theta_{1,2}$ are chosen randomly in the interval $ ]0,2\pi[$ with $\theta_1 \neq \theta_2$ and $\theta_{1,2} \neq \pi$;
    \item The quartic couplings, apart from $\lambda_0$, are chosen randomly in the interval $[-10,10]$;
    \item All scalar squared-masses have to be positive;
    \item The parameters $\mu_1$ and $\lambda_0$ are chosen in such a way that the mass eigenstate $h_4^0$ corresponds to the 125~GeV SM Higgs boson, for which Eq.~(\ref{eq:neutral_high_masses_CP}) provides an excellent approximation;
    \item The state $h_4^0$ has couplings to SM particles - gauge bosons and fermions - at most $10\%$ deviated from the expected SM values.
\end{itemize}
From Eq.~(\ref{eq:neutral_high_masses_CP}), we can extract the following expression for $\lambda_0$:
\begin{equation}
    \lambda_0 \approx \dfrac{m_H^2 - 2 \sqrt{2} \mu_1 u c_\beta \dfrac{s_{\theta_1-\theta_2}}{s_{\theta_2}}}{2 v^2} \, ,
    \label{eq:lambda1_interval1}
\end{equation}
in which we have used $m_{h_4^0}^2 \equiv m_\text{Higgs}^2 = (125~\text{GeV})^2$. From this equation, assuming $\lambda_0$ to be in the real-valued interval $]0,10[$, we get:
\begin{equation}
    0 < m_H^2 - 2 \sqrt{2} \mu_1 u c_\beta \dfrac{s_{\theta_1-\theta_2}}{s_{\theta_2}} < 20 \,v^2 \, .
    \label{eq:mu1_interval1}
\end{equation}
After randomly selecting a value for $\mu_1$ within the regime discussed previously - for which we expect a particle of $\mathcal{O}(v)$ as well as decoupled mass states -, we assure that Eq.~(\ref{eq:mu1_interval1}) is satisfied and the value for $\lambda_0$ is obtained using Eq.~(\ref{eq:lambda1_interval1}). We emphasise that, although we have used our analytical considerations in order to choose the ranges of parameters for numerical scans, all masses and couplings were obtained through the numerical diagonalisation of the exact mass matrices. The selection criteria mentioned indeed give rise not only to a Higgs-like particle but also to positive decoupled squared-masses, namely two neutral, one singly- and one doubly-charged.
\begin{figure}
\centering
\begin{tabular}{cc}
\includegraphics[width=7cm]{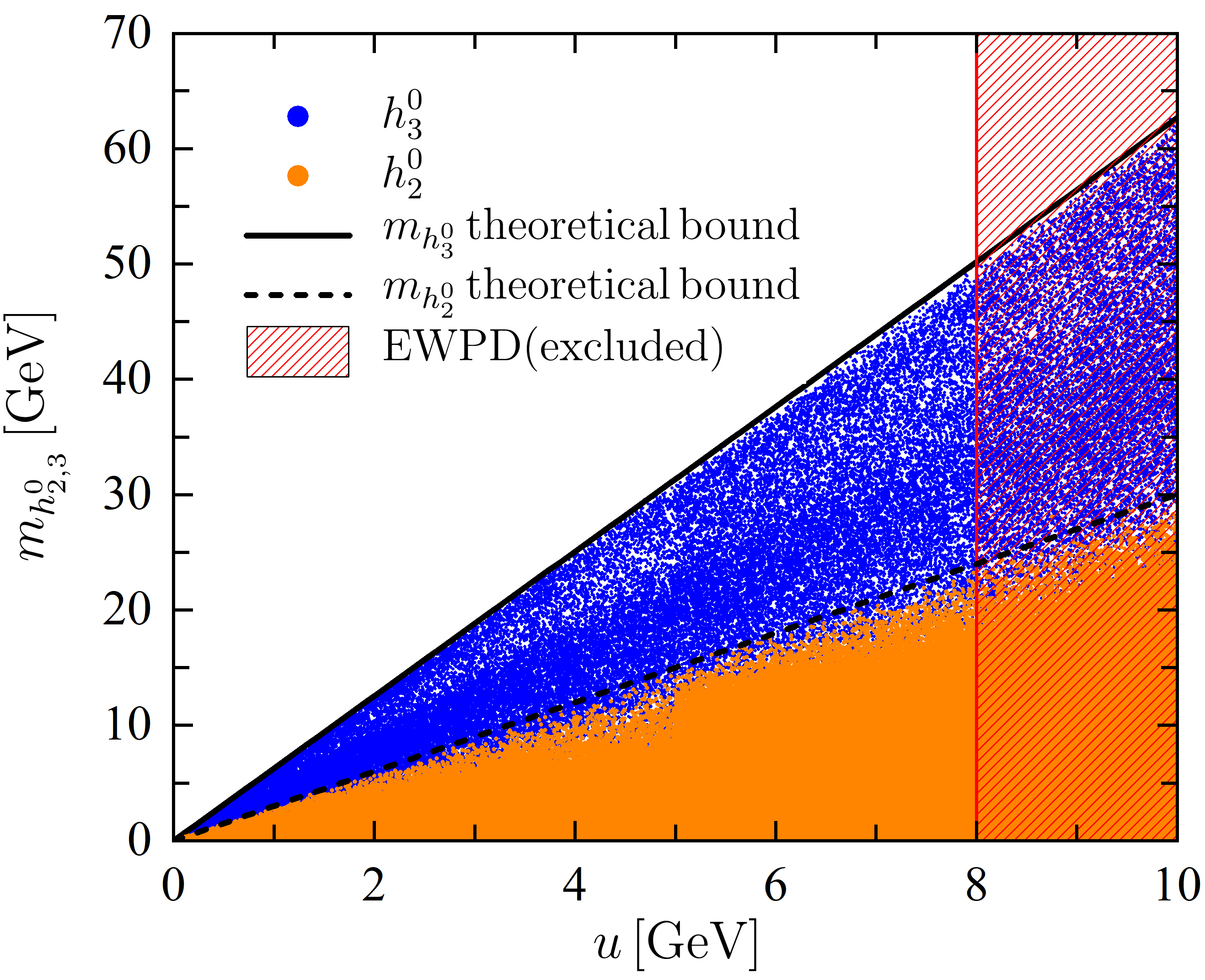} &
\includegraphics[width=7cm]{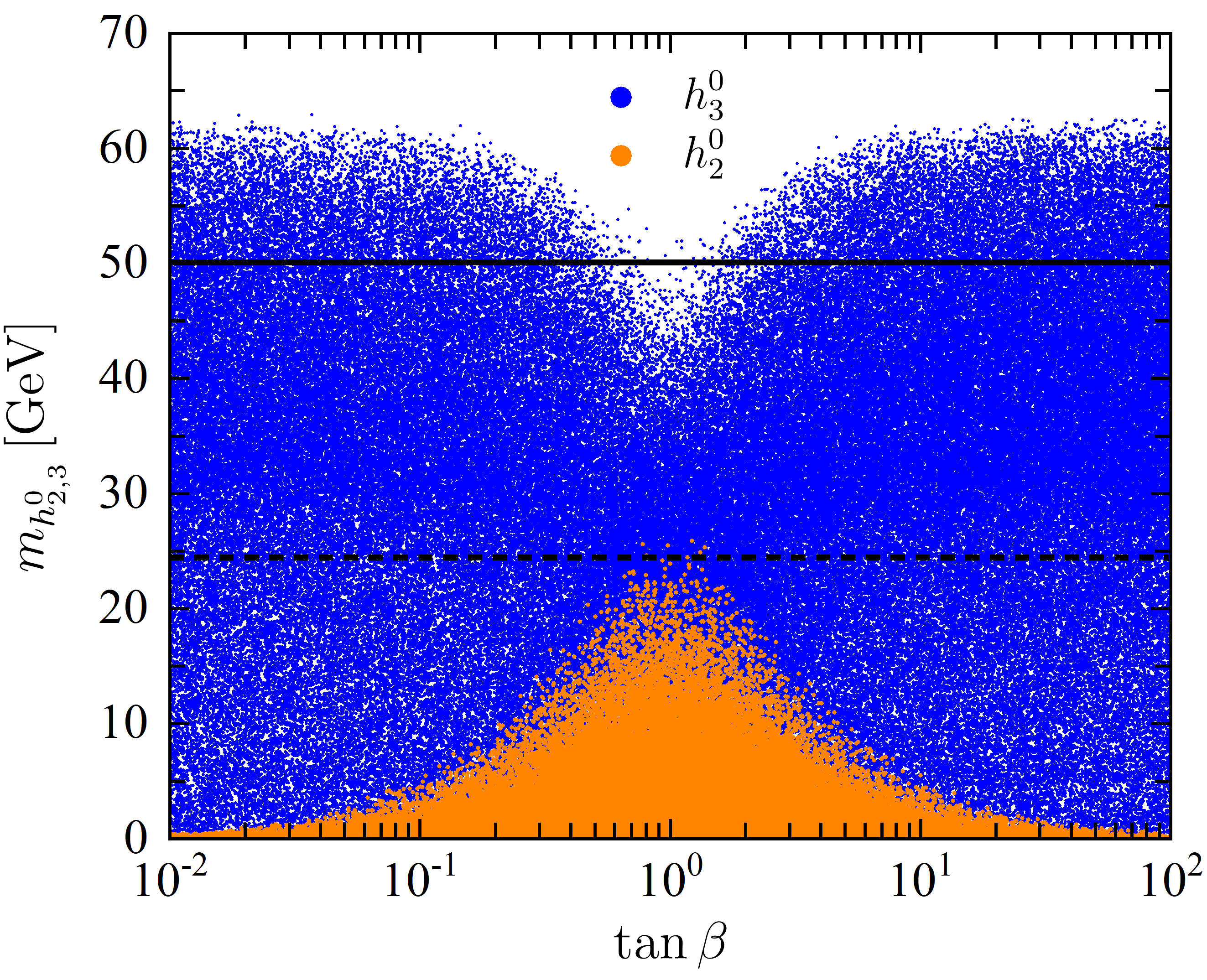}
\end{tabular}
\caption{Left: Scatter plot of the lightest neutral-scalar masses vs. the triplet VEV $u$, in orange and blue for $m_{h_2^0}$ and $m_{h_3^0}$, respectively. The solid (dashed) black line corresponds to the maximum value of the upper bounds given in Eq.~(\ref{eq:CP_lim3}) at each value of $u$. Right: The same as in the left panel but now as function of $\tan\beta$.}
\label{fig:plot1}
\end{figure}
\begin{figure}
\centering
\begin{tabular}{cc}
\includegraphics[width=7.2cm]{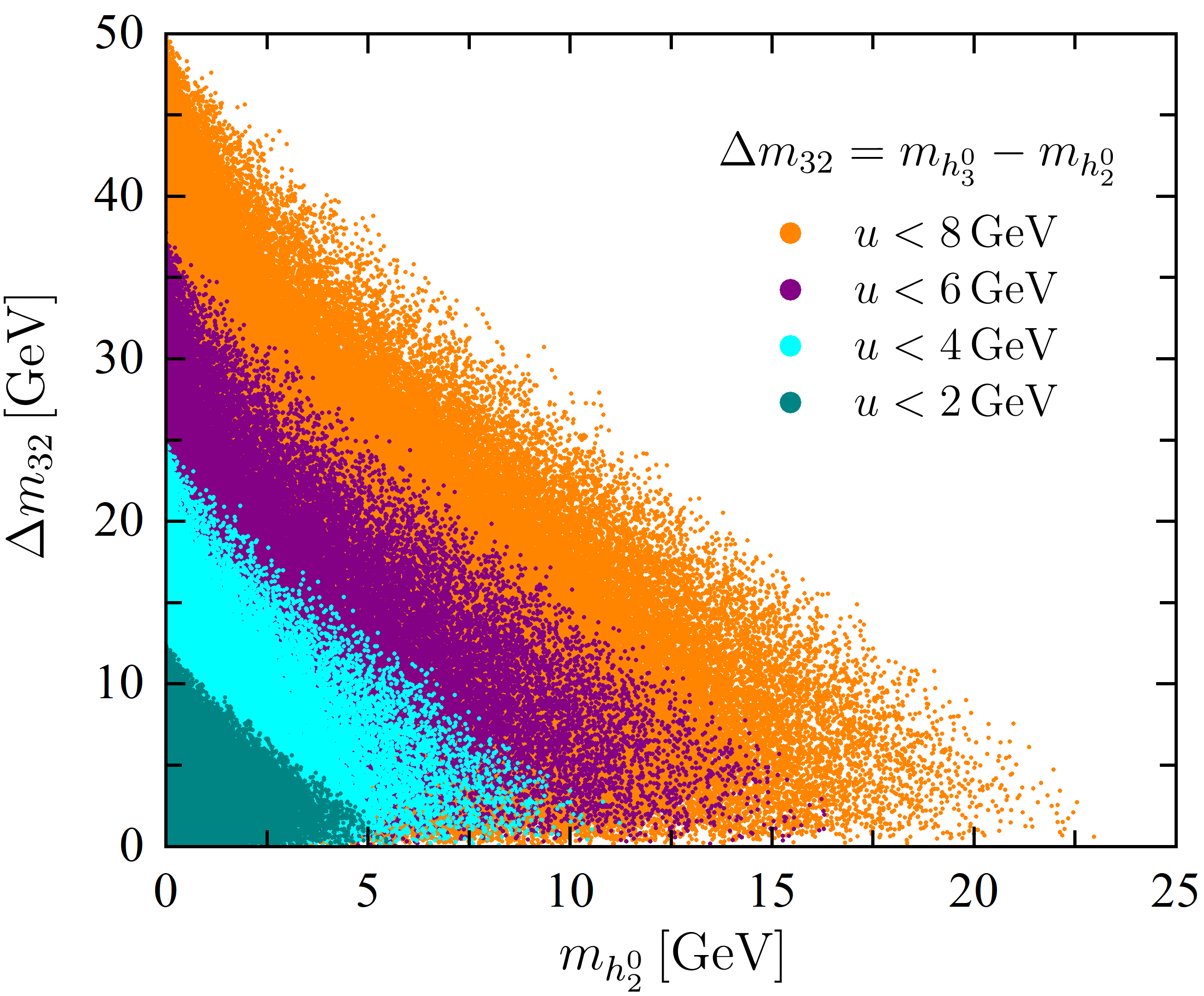} &
\includegraphics[width=7.05cm]{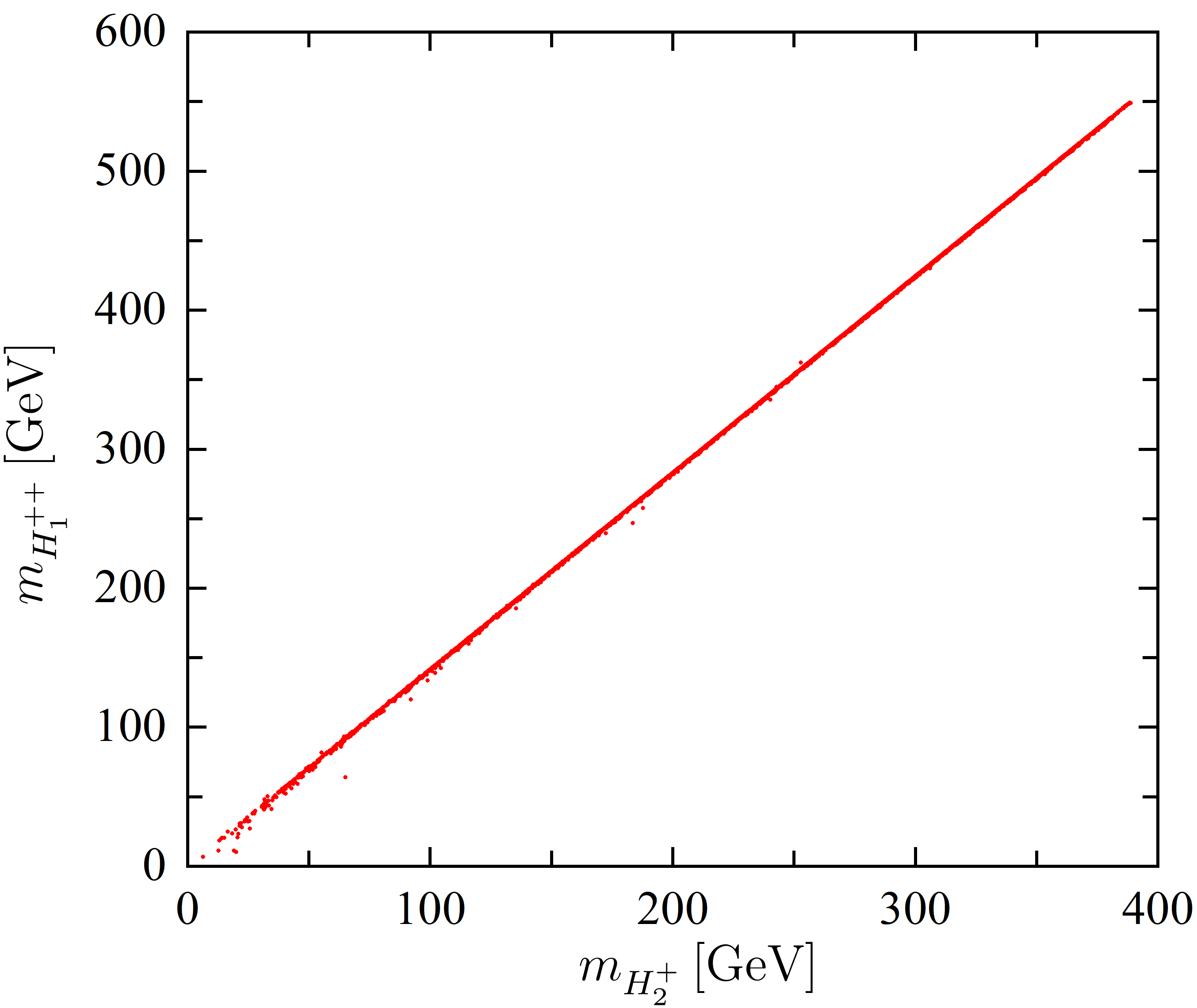}
\end{tabular}
\caption{Left: scatter plots of $\Delta m_{23}$ vs. $m(h_2^0)$ for values of the triplet VEV $u\leq 8,6,4,2$~GeV (in orange, purple, cyan and turquoise, respectively). Right: scatter plot of $m_{H_1^{++}}$ vs. $m_{H_2^+}$ for $u \leq 8$~GeV.}
\label{fig:plot3}
\end{figure}

Under all the above considerations, we have computed the neutral-scalar spectrum for a large sample of the parameter space. Concerning the two lightest scalars $h_2^0$ and $h_3^0$, in the left (right) panel of Fig.~\ref{fig:plot1} we show a scatter plot of their masses against $u$ ($\tan\beta$) in orange and blue for $m_{h_2^0}$ and $m_{h_3^0}$, respectively. The solid (dashed) lines correspond to the theoretical upper bounds found in Eq.~(\ref{eq:CP_lim3}) for $m_{h_3^0}$ ($m_{h_2^0}$). We observe that the numerical scans lead to upper bounds which are in excellent agreement with the approximate analytical prediction. From the right panel, we notice that $\tan{\beta}\simeq 1$ is required to reach $m_{h_2^0}$ near its maximum value. In Fig.~\ref{fig:plot3} we show a scatter plot for the mass difference $\Delta m_{23} \equiv m_{h_3^0} - m_{h_2^0}$ vs. the lightest scalar mass $m_{h_2^0}$ for $u\leq 8,6,4,2$~GeV, in blue, purple, cyan and turquoise, respectively. We see that, as $m_{h_2^0}$ increases, the mass difference $\Delta m_{23}$ decreases, being very small for $m_{h_2^0}$ close to its maximum. Concerning the validity of the approximations deduced in the previous section for the charged-scalar masses, the numerical results plotted in the right panel of Fig.~\ref{fig:plot3} confirm the relation $m_{H_1^{++}} \approx \sqrt{2} \, m_{H_2^{+}}$ stemming from Eq.~\eqref{eq:mcCPV}. This plot also illustrates that, since the quartic parameters $\widehat{\lambda}_{1}$ and $\widehat{\lambda}_{2}$ are limited, those masses are upper bounded. In fact, making once more use of Weyl's inequalities, such bounds can be found. Namely, considering $\text{max}\left( -\widehat{\lambda}_{1},-\widehat{\lambda}_{2} \right)=10$, we get:
%
%
%
\begin{align}
m_{H_2^+} &\lesssim \sqrt{\text{max}\left( - \widehat{\lambda}_{1}/4 , - \widehat{\lambda}_{2}/4 \right)} v \simeq 390 \, \text{GeV} \, ,\\
m_{H_1^{++}} &\lesssim \sqrt{\text{max}\left( - \widehat{\lambda}_{1}/2 , - \widehat{\lambda}_{2}/2 \right)} v \simeq 550 \, \text{GeV} \, ,
\end{align}
which is in very good agreement with the results in Fig.~\ref{fig:plot3} (right panel).

We now address some phenomenological aspects related to those scalars which do not decouple, namely $h_2^0$, $h_3^0$, $H_2^+$ and $H_1^{++}$ (see Table~\ref{tab:final_results}). In particular, we wish to investigate whether such low-mass scalars are at all compatible with existing searches in accelerators. Remember that, although the 125 GeV Higgs boson was discovered at LHC, bounds on its mass and couplings had already been obtained, for instance, at LEP. In spite of the fact that no charged scalars $H^+$ have been detected, measurements of observables such as $b\rightarrow s \gamma$ have been used to impose tight constraints on the mass and couplings of $H^+$. It is therefore of interest to analyse the possibility that, though the SCPV minimum implies light scalars in the 2STM, these new particles might have escaped detection due to their suppressed interactions with SM particles. Consider that, though the extra neutral scalars are the result of mixing with the corresponding neutral fields of the doublet, that mixing is highly suppressed since $u_{1,2} \ll v$. Therefore, since the extra neutral mass eigenstates do have Yukawa interactions with the fermions, we expect them to be highly suppressed.
\begin{figure}
\centering
\includegraphics[width=12cm]{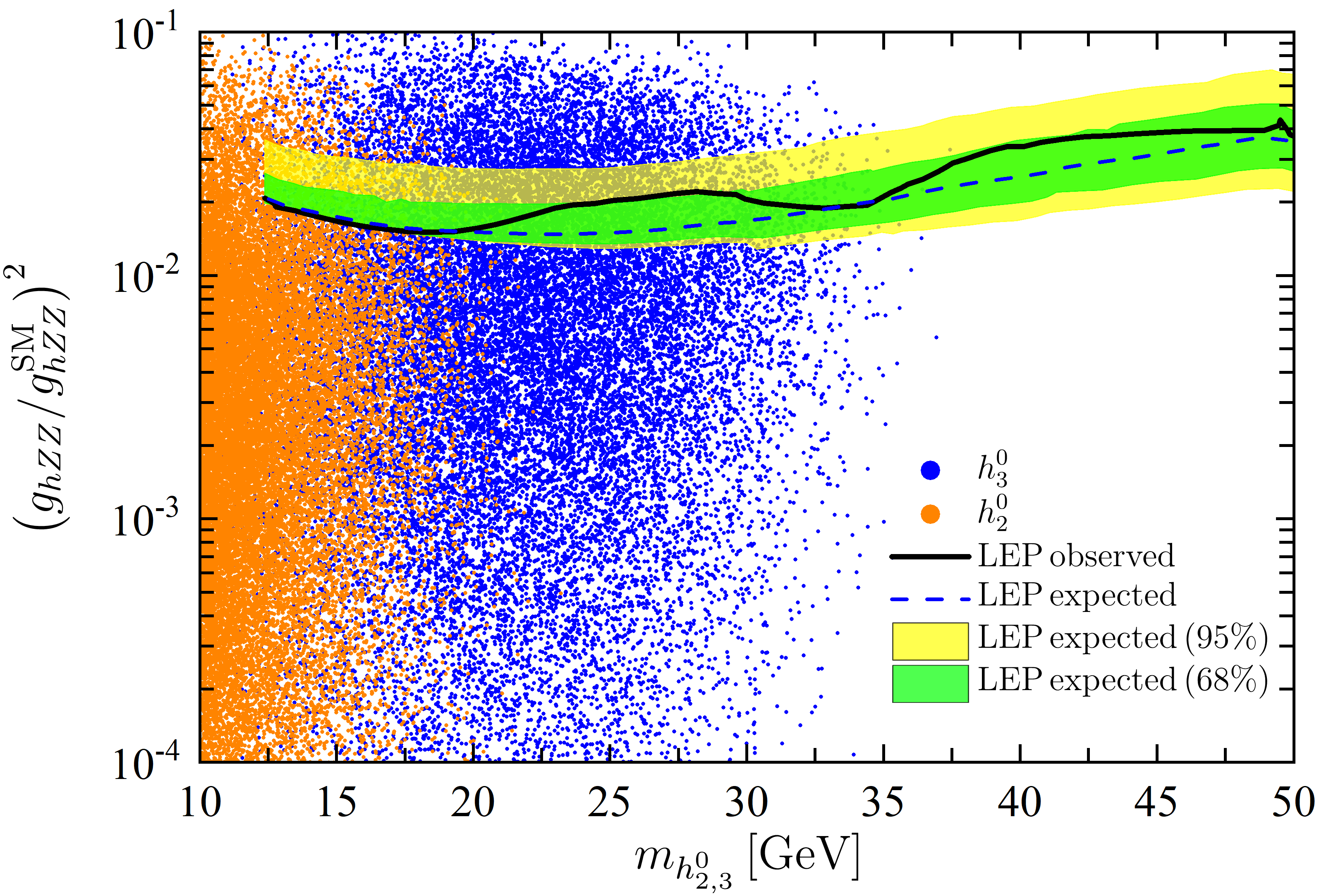}
\caption{Upper bound on the ratio $(g_{hZZ}/g_{hZZ}^{\text{SM}})^2$ at $95 \% $ confidence level, for $m_{h_{2,3}^0}\ge 10~\text{GeV}$, from combined LEP data. The solid black (blue dashed) line stands for the observed (expected) bound. The green (yellow) bands correspond to the $68 \%$~C.L. ($95 \%$~C.L.) regions for the expected bound. In orange and blue we show the $m_{h_2^0}$ and $m_{h_3^0}$ vs. \textit{vs} $(g_{hZZ}/g_{hZZ}^{\text{SM}})^2$ points, for the considered mass range. These results were obtained considering $|u|<8~\text{GeV}$, and the LEP bounds were taken from Ref.~\cite{Barate:2003sz}.}
\label{fig:plot5}
\end{figure}

We start by computing the $hZZ$ couplings for $h\equiv h_{2,3}$. At LEP, the main production process of the Higgs was Higgsstrahlung, $e^+ e^-\rightarrow Z \rightarrow Zh$, whereby a Higgs is emitted by a virtual $Z$ boson. Data from LEP was used to set upper bounds on the $hZZ$-coupling in non-standard models. We numerically compute the ratio $g_{hZZ}/g_{hZZ}^{\text{SM}}$, where $g_{hZZ}$ is the non-standard $hZZ$ coupling and $g_{hZZ}^{\text{SM}}$ the SM one.
Within our numerical scan, such ratio regarding the SM-like Higgs particle must be at most $10~\%$ deviated from the expected SM value. Our results are depicted in Fig.~\ref{fig:plot5} for masses above $10~\text{GeV}$. The fact that $m_{h_2^0}>10 \, \text{GeV}$ also restricts $\Delta m_{32}$, as seen in the left panel of Fig.~\ref{fig:plot3}. This plot shows that $m_{h_2^0}>10$ GeV automatically implies $m_{h_3^0}\lesssim 40$ GeV. This justifies the absence of results near the theoretical upper bound for the lighter
neutral masses for $|u|<8 \, \text{GeV}$. Notice that a large fraction of the  points lie below the upper bounds from LEP searches on low-mass ranges and, thus, such light scalars could have escaped detection at LEP. However, we also see that there is a sizeable portion of the points which are excluded by the same dataset. From now on, we will only consider the parameter space region below the LEP bound.

Turning now to the $H_2^+$ and $H_1^{++}$ charged scalars, we start by looking at the Yukawa couplings of the lightest charged scalar $H_2^\pm$ (recall $H_1^\pm$ is the massless charged Goldstone boson) with the up- ($u$) and down-type ($d$) quarks. In general, these can be written as
\begin{equation}
    - \mathcal{L}_{H_2^\pm} = \dfrac{\sqrt{2} \, V_{ud}^\text{CKM}}{v} X_2 H_2^+ \overline{u} \left( m_d P_R - m_u P_L \right) d \, + \,  \text{H.c.} \quad ,
    \label{eq:chyuk}
\end{equation}
in which $V_{ud}^\text{CKM}$ is the corresponding element of the Cabibbo-Kobayashi-Maskawa (CKM) matrix. Since originally only the doublet $\Phi$ couples to quarks, the charged Yukawa interactions are obtained from the terms featuring $\phi^+$, the upper $\Phi$ component. The real parameter $X_2$ is therefore the fraction of the $H_2^+$ mass eigenstate in $\phi^+$, i.e.
\begin{equation}
    \phi^+ = X_1 H_1^+ + X_2 H_2^+ + X_3 H_3^+ \, .
    \label{eq:phipX}
\end{equation}
\begin{figure}
\centering
\begin{tabular}{cc}
\includegraphics[width=7.1cm]{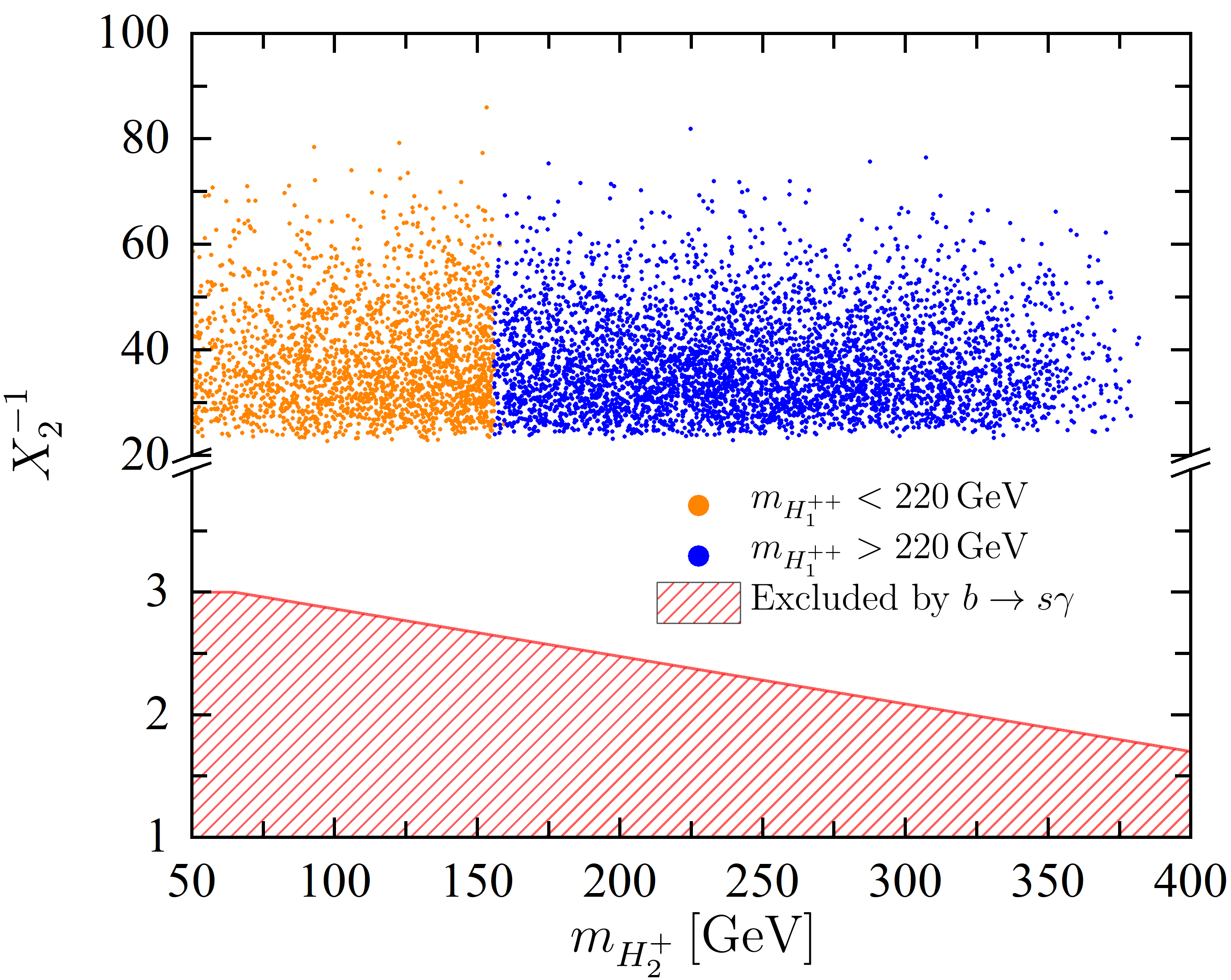} &
\includegraphics[width=6.95cm]{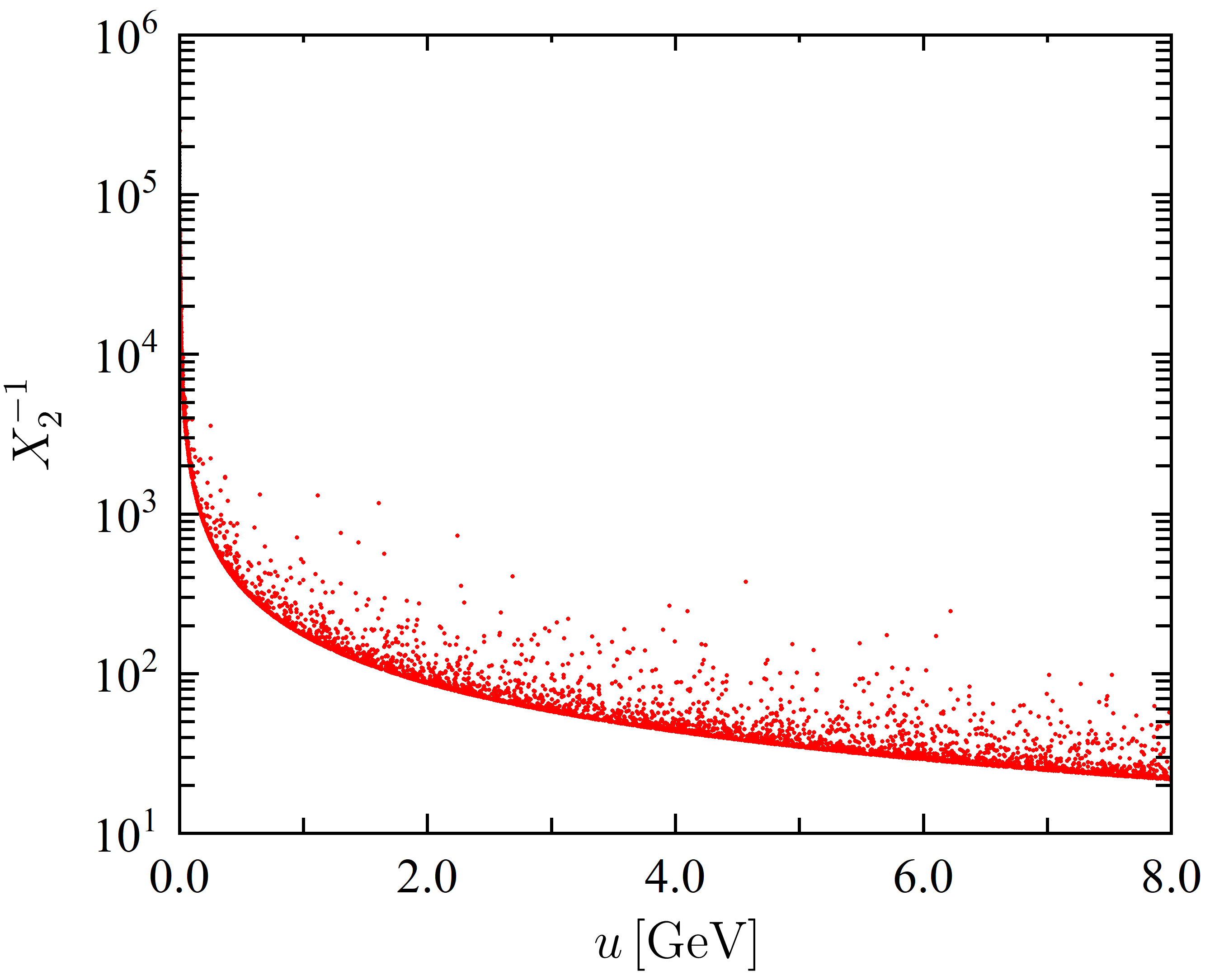}
\end{tabular}
\caption{Left: Allowed points in the $(X_2^{-1},m_{H_2^+})$ plane -- see Eqs.~\eqref{eq:chyuk} and \eqref{eq:phipX}. The hatched region corresponds to the fraction of the parameter space excluded by $b \rightarrow s\gamma$ constraints, as reported in Ref.~\cite{Arbey:2017gmh}. The points for which the lightest doubly-charged scalar mass $m_{H_1^{++}}$ is above (below) the ATLAS lower bound of 220~GeV extracted from $W^\pm W^\pm$ searches are shown in blue (orange) -- see text for more details. Right: $X_2^{-1}$ vs. the triplet VEV parameter $u\lesssim 8$~GeV.}
\centering
\label{fig:plot6}
\end{figure}
The structure of the Yukawa interactions in Eq.~\eqref{eq:chyuk} is identical to those obtained in the type-I 2HDM for the single charged scalar which appears in such model, if one makes the correspondence $X_2 \equiv 1/(\tan\beta)^{\rm 2HDM}$, where $(\tan\beta)^{\rm 2HDM}$ differs from the $\tan\beta$ we have been considering -- $(\tan\beta)^{\rm 2HDM}$ is the ratio of the 2HDM doublet VEVs, whereas in our work $\tan\beta$ is the ratio of triplet VEVs. With this correspondence established, we can use the bounds for the type-I 2HDM charged-scalar interactions obtained in Ref.~~\cite{Arbey:2017gmh}, which are presented as exclusion regions in the $(m_{H^+}, \tan\beta)$ plane. We plot our results in the $( m_{H_2^+},X_2^{-1})$ plane as shown in the left panel of Fig.~\ref{fig:plot6}, where the hatched region originate from the most restrictive bounds coming from $B$ physics, namely from $b \rightarrow s \gamma$ measurements. We thus see that the lightest charged scalar $H_2^+$ is not at all constrained by that data. In 2HDM language, we would say that the whole scanned parameter space corresponds to values of $\tan\beta$ above roughly 20. We also show in blue (orange) the points for which the doubly-charged scalar mass $m_{H_1^{++}}$ is above (below) the experimental lower bound of $220 \, \text{GeV}$ extracted from $W^\pm W^\pm$ ATLAS searches~\cite{Aaboud:2018qcu}\footnote{We do not consider experimental bounds coming from decays of the doubly charged scalars to pairs of same-sign leptons as we have not specified the Yukawa matrices of eq.~\eqref{eq:Lyuk1}. A more detailed study of the model would need to take such bounds into account, as they are typically more stringent than those stemming from the double \textit{W}-boson channel. This would, however, require a detailed analysis of all possible Yukawa interactions between the scalar triplets and the leptons, which goes beyond the scope of the current work. Notice, however, that for $\mathcal{O}$(GeV) triplet VEVs the Yukawa couplings of eq.~\eqref{eq:Lyuk1} required to achieve $\mathcal{O}$(eV) neutrino masses would be so small that the said same-sign dilepton decays would be highly suppressed.}. Using Eq.~\eqref{eq:mcCPV}, such bound corresponds to $m_{H_2^+}>(220\, \text{GeV})/\sqrt{2} \simeq 156 \, \text{GeV}$. From the right panel of Fig.~\ref{fig:plot6}, where $X_2^{-1}$ is plotted against $u$, we see that $X_2^{-1} \gtrsim 20$ stems from the upper limit on $u \lesssim 8$~GeV. Restricting ourselves to neutral scalar masses above 10~GeV, which implies $u \gtrsim 2$~GeV (see Fig.~\ref{fig:plot1}), we can see from the same plot that there is also an upper bound $X_2^{-1} \lesssim 100$.

\section{Conclusions}
\label{sec:conclusions}

Extensions of the SM with $Y=2$ triplet scalar fields are well motivated since small neutrino masses can be
generated via the type II seesaw mechanism. In this work, we investigated the possibility that one might also obtain CP
violation in the leptonic sector via spontaneous symmetry breaking in such models. We have found that a minimum of two
triplet fields are required for SCPV to occur, in a model including a continuous global symmetry, softly broken by three
terms with real coefficients. For comparison, SCPV for a theory invariant under a global continuous symmetry is
not possible
in the 2HDM~\cite{Ferreira:2010hy} or in the STM,  even with generic soft-breaking terms added to the potential. It
is, however, already possible in a 3HDM -- for instance, with a U(1) symmetry under which two of the doublets remain
unchanged. Considering a global U(1) symmetry in the scalar sector allowed us to simplify the potential and, when extended to the fermion sector, it could perhaps generate interesting textures in the neutrino mass matrix. An in-depth analysis of the scalar sector revealed an interesting aspect of the model: when SCPV occurs, the 2STM has no decoupling limit, and two neutral scalars are predicted to have masses lower than the electroweak scale and of the order of the triplet VEVs $u$. We were able to obtain this result using matrix theory theorems to estimate relations between the eigenvalues of the mass matrices, and verified it through a numerical scan of the model's parameter space. This is a result which follows a similar conclusion for SCPV in the 2HDM, as shown in ref.~\cite{Nebot:2020niz} and is due to the fact that the extra minimisation conditions that a CP-violating minimum must obey limit the size of more of the scalar potential's  mass parameters. In the 2STM, however, the existence of non-decoupling when SCPV occurs is also caused by the smallness of the triplets VEVs, necessary for compliance with electroweak precision data. But SCPV is the key ingredient to bring about non-decoupling, since we also verified that CP-conserving minima of the model do not share this non-decoupling regime -- in such a vacuum of the 2STM, even small triplet VEVs do not imply the existence of scalars with masses below the electroweak scale.

We therefore conclude that attempting to simultaneously explain neutrino masses and leptonic CP violation via spontaneous breaking of the CP symmetry in the 2STM would imply the triplets having VEVs of $\mathcal{O}({\rm eV})$, and consequently the existence of two neutral scalars with masses of the same order. Those scalars would have interactions with the known fermions and gauge bosons, and in particular with the neutrinos. The quark interactions should be highly suppressed  due to the fact that they all arise from the Yukawa term in the lagrangean involving the doublet $\Phi$ -- and since the SM Higgs-like state $h^0_4$ is essentially aligned with the neutral, real component of $\Phi$, all quark Yukawa interactions of the lighter scalars, $h^0_{2,3}$, will be suppressed by off-diagonal elements of the rotation matrix between weak and mass scalar eigenstates, which are at least of order $\mathcal{O}(u/v) \sim 10^{-11}$. The gauge interactions of the lighter scalars are not necessarily suppressed, since they will include contributions from the triplet covariant derivative terms. However, we have shown,  through our numerical scans, that the lighter scalars can indeed have interactions with gauge bosons highly suppressed -- our scan had triplet VEVs of order GeV, but the conclusion obviously holds for VEVs of much smaller orders of magnitude. Of course, if these neutral scalars have masses of the eV order, their decays to electroweak gauge bosons or fermions are anyway kinematically forbidden. It would seem, then, that these neutral scalars could be thought of as possible Dark Matter  candidates. But in a reversal of the usual reasoning, one would expect that the decays to neutrinos of these lighter scalars would be relevant. In fact, for the neutrino mass explanation arising in this model to be ``natural", one would expect the scalar-neutrino Yukawa couplings to be of order $\mathcal{O}(1)$, which would therefore make the decay channel $h^0_{2,3}\rightarrow \nu\bar{\nu}$ be significant, maybe even dominant, precluding their stability, and thus invalidating their validity as Dark Matter candidates.

The SCPV 2STM could however be a valid description of particle physics if the triplet VEVs, while being small enough so that the model complies with electroweak precision data, are of GeV order. A numerical scan of the model then shows that there are  regions of parameter space for which the lighter scalars would have evaded detection at LEP; and for which the charged and doubly charged scalars would have masses and couplings such that they would not yet have been discovered. This raises the possibility that lighter scalars than the 125 GeV Higgs boson may be discovered at LHC or elsewhere -- one such possibility (though not possible in this model, due to the masses and couplings involved for the lighter scalars) would be the possible excess at 96 GeV observed by CMS~\cite{Sirunyan:2018aui}, which can be interpreted in several ways in BSM theories~\cite{Fox:2017uwr,Biekotter:2017xmf,Haisch:2017gql,Liu:2018ryo,Biekotter:2019kde,AguilarSaavedra:2020wmg}. A full study of the phenomenology of this model is therefore of interest to fully ascertain the likelyhood of its being fully probed at LHC or future colliders.



\acknowledgments
This work is supported by Funda\c{c}{\~a}o para a Ci{\^e}ncia e a Tecnologia (FCT, Portugal) through the projects UIDB/00777/2020, UIDP/00777/2020, UIDB/00618/2020, CERN/FIS-PAR/
0004/2019, CERN/FISPAR/0014/2019, PTDC/FIS-PAR/29436/2017 and by HARMONIA project's contract UMO-2015/18/M/ST2/00518. The work of B.L.G. is supported by the FCT grant SFRH/BD/139165/2018.

\end{document}